\documentclass[letterpaper,twocolumn,10pt]{article}
\usepackage{usenix-2020-09}
\microtypecontext{spacing=nonfrench} 

\usepackage{xspace}
\usepackage[normalem]{ulem}

\usepackage{enumitem}

\usepackage{tabularx}
\usepackage{multirow}
\usepackage{subcaption}
\usepackage{color}
\usepackage{booktabs}
\usepackage{appendix}

\usepackage{cite}
\usepackage{amsmath,amssymb,amsfonts,amsthm}
\usepackage{mathtools}
\usepackage{algorithm}
\usepackage{algpseudocode}
\usepackage{graphicx}
\usepackage[locale=US,binary-units=true]{siunitx}

\graphicspath{{./Figures/}}
\DeclareGraphicsExtensions{.pdf,.jpeg,.png}
\usepackage{cleveref}
\crefname{section}{§}{§§}
\Crefname{figure}{Figure}{Fig.}
\crefname{figure}{Figure}{Fig.}
\crefname{appendix}{Appendix}{App.}
\crefname{table}{Table}{Tbl.}
\Crefname{equation}{Eq.}{Eq.}
\crefname{equation}{Eq.}{Eq.}

\usepackage{xcolor}
\definecolor{orange}{RGB}{255,153,51}
\definecolor{lightorange}{RGB}{255,235,214}
\definecolor{darkorange}{RGB}{171,68,1}
\definecolor{darkerorange}{RGB}{171,68,1}

\definecolor{lightgrey}{RGB}{242,242,242}
\definecolor{midgrey}{RGB}{191,191,191}
\definecolor{darkgrey}{RGB}{90,90,90}

\definecolor{darkblue}{RGB}{29,117,180}
\definecolor{lightblue}{RGB}{205,224,220}

\definecolor{darkred}{RGB}{255,32,33}
\definecolor{lightred}{RGB}{255,128,129}

\definecolor{darkgreen}{RGB}{101,178,50}
\definecolor{lightgreen}{RGB}{170,223,135}

\definecolor{cLightRed}{HTML}{E74C3C}
\definecolor{cRed}{HTML}{C0392B}
\definecolor{cBlue}{HTML}{2980B9}
\definecolor{cLightBlue}{HTML}{3498DB}
\definecolor{cDarkBlue}{HTML}{10334A}
\definecolor{cGreen}{HTML}{27AE60}
\definecolor{cLightGreen}{HTML}{2ECC71}
\definecolor{cViolet}{HTML}{8E44AD}
\definecolor{cLightViolet}{HTML}{9B59B6}
\definecolor{cOrange}{HTML}{D35400}
\definecolor{cLightOrange}{HTML}{E67E22}
\definecolor{cYellow}{HTML}{F39C12}
\definecolor{cLightYellow}{HTML}{F1C40F}
\definecolor{cGray}{HTML}{909090}

\usepackage{tikz}
\usepackage{xfp}
\usepackage{pgfplots}

\usetikzlibrary{calc, patterns, math, plotmarks}
\usepgfplotslibrary{fillbetween, groupplots}

\pgfplotsset{
    every axis/.append style={
      label style={font=\normalsize},
      tick label style={font=\normalsize}  ,
      every x tick/.style={color=black, thin},
      every y tick/.style={color=black, thin},
    },
    every axis plot/.append style={line width=1pt},
}

\makeatletter
\g@addto@macro{\UrlBreaks}{\UrlOrds}
\makeatother

\newcommand{\myitem}[1]{\vspace{0.01in}\noindent\textbf{#1}}

\newcommand{\system}[0]{Oscilloscope\xspace} 

\newcommand{\comment}[1]{}
\newcommand{\remove}[1]{}

\newcommand{\ms}{\ensuremath{\,\text{ms}}\xspace}
\newcommand{\us}{\ensuremath{\,\mu\text{s}}\xspace}

\newcommand{\eg}{\textsl{e.g.},\xspace}
\newcommand{\ie}{\textsl{i.e.},\xspace}

\begin{document}

\date{}

\title{\Large \bf \system: Detecting BGP Hijacks in the Data Plane}

\author{%
{Tobias B\"uhler}\\
ETH Z\"urich\\
\url{buehlert@ethz.ch}
\and
{Alexandros Milolidakis}\\
KTH Royal Institute of Technology\\
\url{miloli@kth.se}
\and
{Romain Jacob}\\
ETH Z\"urich\\
\url{jacobr@ethz.ch}
\and
{Marco Chiesa}\\
KTH Royal Institute of Technology\\
\url{mchiesa@kth.se}
\and
{Stefano Vissicchio}\\
University College London (UCL)\\
\url{s.vissicchio@ucl.ac.uk}
\and
{Laurent Vanbever}\\
ETH Z\"urich\\
\url{lvanbever@ethz.ch}
} 

\maketitle

\begin{abstract}

The lack of security of the Internet routing protocol (BGP) has allowed
attackers to divert Internet traffic and consequently perpetrate service
disruptions, monetary frauds, and even citizen surveillance for \emph{decades}.
State-of-the-art defenses rely on geo-distributed BGP monitors to detect rogue
BGP announcements. As we show, though, attackers can easily evade detection by
engineering their announcements.

This paper presents \system, an approach to accurately detect BGP hijacks by
relying on real-time traffic analysis. As hijacks inevitably change the
characteristics of the diverted traffic, the key idea is to track these changes
in real time and flag them. The main challenge is that ``normal'' Internet
events (e.g., network reconfigurations, link failures, load balancing)
\emph{also} change the underlying traffic characteristics -- and they are
\emph{way} more frequent than hijacks. Naive traffic analyses would hence lead
to too many false positives.

We observe that hijacks typically target a \emph{subset} of the prefixes
announced by Internet service providers and only divert a subset of their
traffic. In contrast, normal events lead to more uniform changes across prefixes
and traffic. \system uses this observation to filter out non-hijack events by
checking whether they affect multiple related prefixes or not.

Our experimental evaluation demonstrates that \system quickly and accurately
detects hijacks in realistic traffic traces containing hundreds of events.

\end{abstract}

\section{Introduction}
\label{sec:introduction}
BGP \textit{hijacks} have plagued the Internet over the last decade, targeting
critical societal services (\eg finance services)~\cite{financial-hijack}, the
core of the Internet infrastructure (\eg the DNS system)~\cite{dns-hijack}, and
citizens (\eg for surveillance purposes)~\cite{citizens-surveillance-hijack}.
Even worse, several BGP hijacks have gone unnoticed for a long time~\cite{3ves}.
The core problem is the lack of in-built security mechanisms in the interdomain
routing protocol, \ie the Border Gateway Protocol (BGP), which makes launching a
BGP hijack attack a simple task. Any network can falsely announce in BGP as the
legitimate owner of any IP prefix, and all other networks will trust such an
announcement. So, when the attacker's BGP announcement propagates in the
Internet, some networks will reroute their traffic towards the attacker's
network.

Defenses to BGP hijacks aim to prevent them entirely or detect them reliably.
Existing solutions to prevent hijacks incorporate crypto operations into BGP to
verify the exchanged messages. However, complete prevention is tough to achieve.
For example, BGPSec~\cite{lepinski2017bgpsec} faces insurmountable deployability
barriers, while RPKI~\cite{rpki} only protects against limited hijack types.
Proposals to detect hijacks, in contrast, are often easy to deploy but have
other shortcomings. Control-plane-based solutions are inherently limited by the
visibility of their vantage points and cannot detect all attacks (as our
analysis shows in \cref{sec:motivation}). Data-plane-based systems
(\eg~\cite{ispy}) often require a large number of active probes, which leads to
scalability problems, or they can only detect specific hijack types (\eg
blackholes).

In this paper, we introduce \system, a system that detects BGP hijacks by
observing and comparing hijack-specific characteristics in passively collected
data-plane traffic. \system is designed to run standalone, analyzing the
\textit{existing} traffic crossing the network it is deployed in.

\system's design leverages two simple yet powerful observations: \textit{(i)}\,
the existing data-plane data already carries ``signals'' which can be linked to
an ongoing hijack attack (e.g., also shown in~\cite{dart,blackhat}); and
\textit{(ii)}\, the often-used per-neighbor routing
policies~\cite{muhlbauer2007search} result in forwarding paths that are similar
(or equal) for traffic flows between any pair of networks (\ie ASes) in the
Internet. More concretely, the first observation implies that when an attacker
intercepts a set of flows towards one prefix, the packets' Round Trip Times
(RTTs) in the diverted traffic flow will increase (or decrease). In general, an
interception attack fundamentally forces a change in the traffic forwarding path
that cannot be avoided or hidden by the attacker. The second observation implies
that ``normal'' forwarding changes (\eg due to a link failure) will equally
affect all the traffic flows from \system's network and a specific external AS.
In contrast, a hijack targeting a small set of our prefixes will break this
similarity and build an exploitable comparison point.

\begin{figure}
	\centering
	\includegraphics[width=1\columnwidth]{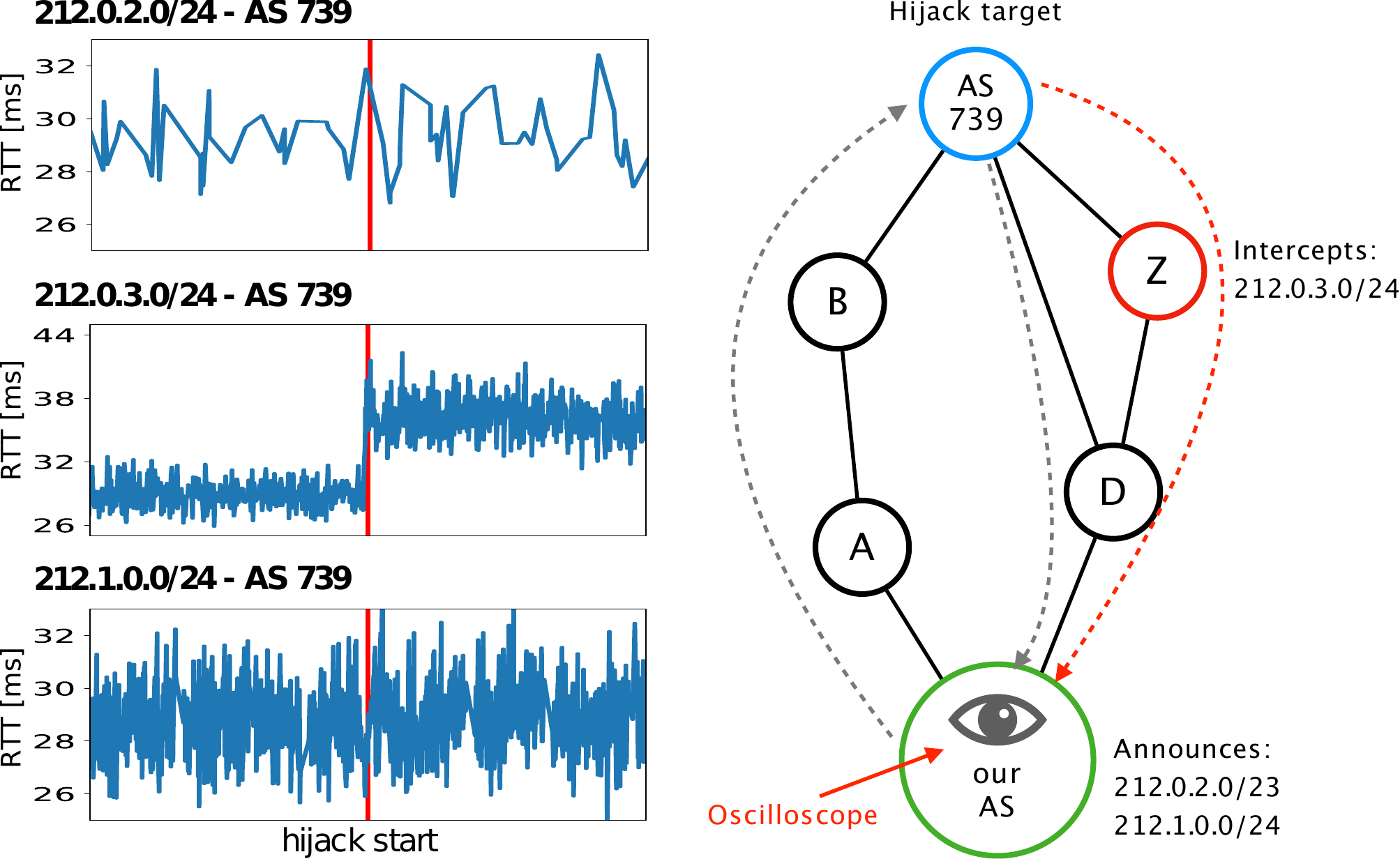}
    \caption{As Z intercepts traffic from AS 739 towards our prefix
    212.0.3.0/24, which leads to an observable RTT change in the corresponding
    ``combined signal''. Note that traffic between other prefixes and AS 739
    does not show the change that \system uses as comparison points.}
	\label{fig:hijack_example}
\end{figure}

\system detects hijacks in three steps, illustrated in the example in
\cref{fig:hijack_example}. In the example, AS Z intercepts traffic towards
the prefix 212.0.3.0/24, which is announced by our AS where \system is deployed.

First, \system combines collected RTT samples to build so-called ``combined
signals,'' which contain all the RTT samples from flows that share the same
local /24 prefix and target destinations belonging to one external AS (the
figure shows three examples). Aggregating measurements this way improves the
practicality and scalability of \system, in terms of both data to maintain and
comparisons to perform. We do not lose any information, given that /24 is the
smallest prefix size globally advertised by BGP~\cite{min_24}, and hence hijacks
cannot target prefixes more specific than /24s.

Second, according to our insight \textit{(i)}, \system detects changes in the
combined signals. It compares the long-term minimum RTT value against RTT values
in consecutive short-term windows and flags changes bigger than a given
threshold (\eg see the middle plot in \cref{fig:hijack_example}).
Comparing the RTT across consecutive short-term values filters out noise. Since
the minimum RTT can be computed efficiently, changes for all combined signals
are detected in near real-time.

Finally, \system performs two-sample statistical tests between RTT samples from
the combined signal, which observes the change and its ``buddy'' prefixes (\ie
the combined signals not affected by the hijack in \cref{fig:hijack_example}) --
where buddy prefixes are defined based on our observation \textit{(ii)}. In the
case of a hijack-induced, prefix-specific change, the test will indicate that
the samples come from different distributions increasing \system's hijack
evidence.

We implemented \system and used it to analyze hundreds of network events. We show
that Oscilloscope reliably detects blackhole and interception attacks (recall of
94\%) while ignoring most ``normal'' forwarding events (precision of 93\%) found
in more than six hours of traffic traces collected in a realistic virtual
environment. \system detects most events in fewer than 10 seconds.

\myitem{Contributions.} We make the following contributions:

\begin{enumerate}[nosep,leftmargin=*]
  \item We perform studies to show that interception attacks (invisible to
  current control-plane-driven detection systems) are easy to compute, have a
  good probability of succeeding, and can affect many networks
  (\cref{sec:related_work});
  \item We introduce \system which can detect interception and blackhole attacks
  using data-plane traffic (\cref{sec:overview});
  \item We explain how \system can detect changes in noisy RTT signals
  (\cref{sec:detection}) and validates its observations
  (\cref{sec:validation});
  \item We perform an in-depth analysis of \system's performance
  (\cref{sec:evaluation}) based on RTT signals from simulated hijack events in
  a realistic virtual environment (\cref{sec:exp});
  \item We confirm that \system's requirements are fulfilled in the real
  Internet based on control-plane data (\cref{sec:deployment}).
\end{enumerate}

\section{Background, Related Work, and Motivation}
\label{sec:related_work}

In this section, we introduce the relevant background and our attacker model.
Afterwards, we discuss related work and how hijackers can evade monitor-based
detection systems.

\subsection{Background and attacker model}
\label{sec:background}

\myitem{Basics.}
The Border Gateway Protocol (BGP \cite{bgp}) is the de-facto routing protocol
that glues together all the tens of thousands of Autonomous Systems (ASes)
forming the Internet. Using BGP, neighboring ASes generate, receive, and
propagate announcements about the \textit{routes} that can be used to reach the
announced IP prefixes. Each AS is identified by an \textit{AS number}, and each
route contains an \textit{AS path} field, which contains a sequence of AS
numbers representing the ASes through which an announcement has propagated. Each
BGP router within an AS selects a single best route towards each IP prefix and
processes each prefix independently. A router selects routes based on
operator-defined routing policies and attributes such as the AS path length.

For instance, an AS will likely prefer to route traffic destined to an IP prefix
through one of its customers (which pays money) rather than using a provider
(which costs money) regardless of the AS path length~\cite{gill2014survey}.
Similarly, an AS may not be willing to exchange traffic between two providers as
it does not gain any monetary benefits.

\myitem{Hijack attacks.} BGP has not been designed with security in mind
\cite{butler2009survey}. Any network may announce an IP prefix regardless of
whether they lawfully own that prefix or not, a so-called BGP \textit{hijack}
attack. Since other networks do not have the means to verify the legitimacy of
the information contained in a BGP announcement, networks may inadvertently
select malicious routes and reroute their traffic towards the adversarial
networks. We say that a hijack attack is a blackhole hijack if the malicious
network cannot forward the traffic to the legitimate destination after hijacking
it. For example, if the hijack also affects all the neighbors of the malicious
network. In this case, the traffic is dropped. We refer instead to an
interception hijack if the malicious network can still forward the traffic to
the legitimate destination (i.e., a neighbor of the malicious network is still
selecting a legitimate route).

\myitem{Attacker model.} We assume that an attacker either performs
\textit{same-prefix} attacks, \ie the attacker announces the same IP prefix and
length as the legitimate prefix; or \textit{more-specific} attacks, \ie the
attacker announces a more-specific prefix than the legitimate one. Note that
same-prefix attacks do not propagate through the entire Internet. Some regions
of the Internet will select the legitimate announcement, and others will choose
the illegitimate one, making detecting these attacks more challenging. In both
cases, the hijacker might forward the attracted traffic back to the victim AS if
a return path is available or through a pre-configured tunnel (which in both
cases results in an interception hijack) or drop the collected traffic
(resulting in a blackhole hijack). To achieve that, the attacker can use
techniques such as AS path poisoning~\cite{poisoning} (\ie adding specific AS
numbers to the announced AS path) to limit the spread of its advertisements or
keep a return path open for an interception attack (when traffic tunneling
cannot be used)~\cite{goldberg2010secure}.

\begin{figure}
	\centering
	\includegraphics[width=1\columnwidth]{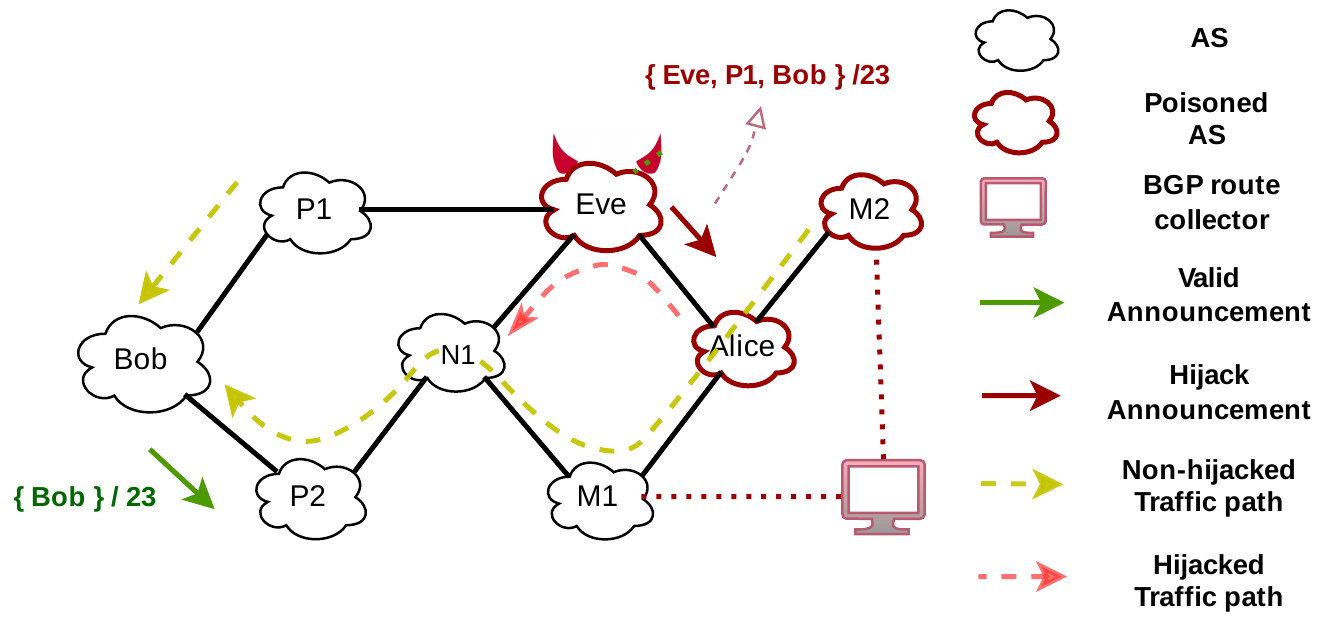}
	\caption{Hijack example of Eve intercepting Bob's traffic from Alice. M1 and
	M2 are the monitoring ASes that report their BGP routes to the BGP
	collector.}
	\label{fig:hijack}
\end{figure}

\myitem{Hijack example.}
\cref{fig:hijack} shows an example of a same-prefix interception attack. All
networks select their best BGP routes based on the shortest AS-path length. For
this example, we do \emph{not} consider that M1 and M2 act as BGP monitors. Bob
has two providers, P1 and P2, and configures BGP to announce its /23 IP prefix
to both of them. P1 has a neighbor Eve and does not announce routes to Eve. N1
has two providers, Eve and M1, and it decides to announce Bob's route only to
M1. Bob's announcement, therefore, propagates through Alice and M2. (Alice may
propagate the announcement to Eve, but this is irrelevant in this example). Now,
imagine Eve wants to intercept the traffic flow from Alice to Bob. Eve can
generate a bogus BGP announcement towards Alice, stating that it has a route
towards Bob's network for the same IP prefix that goes through P1. Since this
route has only three hops (\ie Eve, P1, and Bob) while Alice's currently
selected route has four hops (\ie M1, N1, P2, and Bob), Alice will select this
new route towards Eve and divert her traffic accordingly. Eve can now either
drop the traffic (\ie a blackhole attack) or reroute the traffic to any of its
neighbors that still have a valid route towards the legitimate Bob network, \eg
N1, which results in an interception attack.\footnote{N1 could easily detect
that it is receiving traffic for a non-announced prefix. However, today's
networks do \emph{not} check the consistency of the BGP routes (control plane)
with data-plane observations.}

In a more-specific attack, Eve would announce a /24 subnet of the original /23
prefix originated by Bob. Since BGP treats different IP prefixes independently,
Eve's announcement propagates everywhere. Eve can additionally ``poison'' the AS
path by including P1 and Bob. As a result, the BGP loop detection
algorithm~\cite{bgp} running on P1's and Bob's routers will drop the /24 route.
They continue to use the original /23 route.

\myitem{More-specific vs. same-prefix attacks.} Limiting the spread of
more-specific attacks is more challenging than for same-prefix attacks since
their propagations are inherently limited by the competing, legitimate
announcement, which may be preferred in some parts of the Internet (\eg when the
AS path is shorter). More-specific hijacks do not compete against any other
legitimate announcement, so they tend to propagate everywhere (unless one
carefully exploits AS path poisoning).

\subsection{Analysis of existing solutions}
\label{sec:Solutions}

\myitem{Secure routing protocols are either undeployable or have limited
effectiveness.} Over the years, multiple techniques and modifications to BGP
have been proposed to \textit{proactively} secure the Internet against BGP
hijacking~\cite{butler2009survey, testart2018reviewing}. However, the lack of
global consensus, the ever-increasing size of the BGP tables~\cite{BGP2019}, as
well as the extra overhead required to authenticate each entry have led to
complex cryptographic signing techniques~\cite{kent2000secure,
van2007interdomain, white2003securing}, such as
BGPSec~\cite{lepinski2017bgpsec}, to be far from deployable in
production~\cite{zhao2005performance, testart2018reviewing}. As such, simpler
forms of crypto-based mitigations have been proposed. For instance,
RPKI~\cite{rpki} only stores a digitally-signed mapping between an IP prefix and
the AS number allowed to originate a route for it. Unfortunately, previous
results have shown that by picking a victim and malicious ASes at random over
the Internet, the probability of attracting at least 10\% of the traffic, even
with RPKI fully deployed, was above 60\%~\cite{goldberg2010secure}.

\myitem{Monitor-based countermeasures have limited visibility.}
To overcome the limited deployment and effectiveness of crypto-based solutions,
today's networks rely on \textit{reactive} approaches that monitor the BGP
announcements propagated through the Internet from various vantage
points~\cite{artemis, argus, buddyguard}. These systems typically provide
real-time detection within \SI{5}{s} by analyzing BGP announcements through the
publicly available BGP monitoring infrastructure, \ie RIPE RIS~\cite{Ripe-ris},
BGPStream~\cite{bgpstream}. An inherent limitation of such solutions is that
same-prefix BGP hijacks are only detectable if any monitor receives illegitimate
announcements. We show in \cref{sec:motivation} that monitor-based detection
systems can easily be evaded by carefully crafting BGP announcements that do not
propagate to the monitors.

\myitem{Probing-based data-plane countermeasures can be evaded.}
Data-plane approaches rely on active probing measurements (e.g., pings,
traceroutes, nmap, ...) often performed between external probe locations and the
victim's network. Since traffic is forwarded to the attacker during the hijack
period, we may discover hijacks by probes that fail to reach their destinations.
Those approaches' success heavily relies on their ability to distinguish hijack
signals from common traceroute problems. As the hijacker can attack any
sub-prefix of the victim's prefix, detecting all possible attack scenarios with
low false positive rates (or false positives in case of interceptions) is
time-consuming and computationally expensive.

For example, iSPY~\cite{ispy} offers a lightweight approach with low false
positives that detect hijacks by observing unusual changes to usually stable
traceroute paths. However, it protects only against blackholing and same-prefix
attacks. Zheng et Al.~\cite{zheng2007light} discover interception attacks by
observing AS hop count changes in traceroutes. However, an interception hijacker
can easily manipulate this information by replying with false IPs. Hu et
Al.~\cite{hu2007accurate} extract ``fingerprints'' from the devices of the
correct origin and compare them with the fingerprints of the potential hijacker.
Since the hijacker cannot completely mimic and lie about a network, blackholes
become distinguishable, yet interceptions easily evade detection as the traffic
reaches the correct origin. As such, systems operating in the data plane have
been challenging to commercialize.

\myitem{Naive RTT-based solutions generate false positives or are challenging to
deploy.} There exists other work which uses RTT measurements to detect certain
BGP hijacks. The RTT-based detection method in~\cite{hijack-rtt-student,dart}
uses a simple, threshold-based change-detection algorithm to observe possible
attacks. \cite{crowd} focuses on a crowd-based detection approach that
highlights possible routing anomalies. Both methods have difficulties in
distinguishing hijacks from other events. \system can validate some of its
observations using statistical tests and buddy prefixes.
DARSHANA~\cite{darshana} also uses changes in the RTT to detect hijacks but
actively generates RTT samples with a new ``cryptographic'' ping protocol which
requires the exchange of public keys with the destination. In comparison,
\system does not need support from other networks to detect possible hijacks.

\subsection{Evading monitor-based detection}
\label{sec:motivation}

We now show how hijackers can evade detection from the state-of-the-art
monitor-based detection system Artemis~\cite{artemis}. The main idea is to
generate interception attacks that do not disrupt the hijacked traffic so that
the malicious BGP announcements do not propagate to the BGP monitors. All the
attacks are RPKI valid by consistently placing the legitimate owner as the
originator of an announcement.

Similarly to~\cite{artemis}, we rely on a large-scale BGP simulator to model the
propagation of BGP announcements Internet-wide. The simulator is based on CAIDA
topologies~\cite{CaidaASrelations}, augmented with IXP datasets~\cite{CAIDAIXP},
and the routing policies follow the standard Gao-Rexford
conditions~\cite{gao2001stable}: prefer customer routes over peers and provider
ones and do not export provider-to-provider routes. See \cref{app:Simulator}
and~\cite{hijack-poster} for more details.

\myitem{A stealthy hijacker.} We introduce a hijack attacker that crafts
malicious BGP announcements to avoid propagating to any BGP monitors using AS
path poisoning attacks~\cite{poisoning}. Concretely, the attacker computes a set
of ASes to include in the AS path of the bogus announcements so that the route
is both \textit{i)} long enough to avoid propagating far in the Internet and
\textit{ii)} contains carefully selected AS numbers of critical ASes (\eg
transit ASes). The BGP loop-detection mechanism~\cite{bgp} prevents these ASes
from propagating the bogus route further. For instance, in \cref{fig:hijack}, M1
and M2 are networks that send their BGP routes to a BGP route collector (\ie BGP
monitors). Eve may announce Bob's IP prefix with the following AS path <Eve, M2,
Bob>. In this case, Alice selects this three-hop route which does not propagate
to M2. Also, M1 still selects the <N1, P2, Bob> route, which is shorter than
<N2, Alice, Eve, M2, Bob>. The attack is not visible from the monitors.
\cref{app:stealthy-hijacker} gives more details.

\begin{figure}%
	\centering
	\begin{subfigure}[t]{0.5\linewidth}
		\vskip 0pt 
		\includegraphics[width=\linewidth]{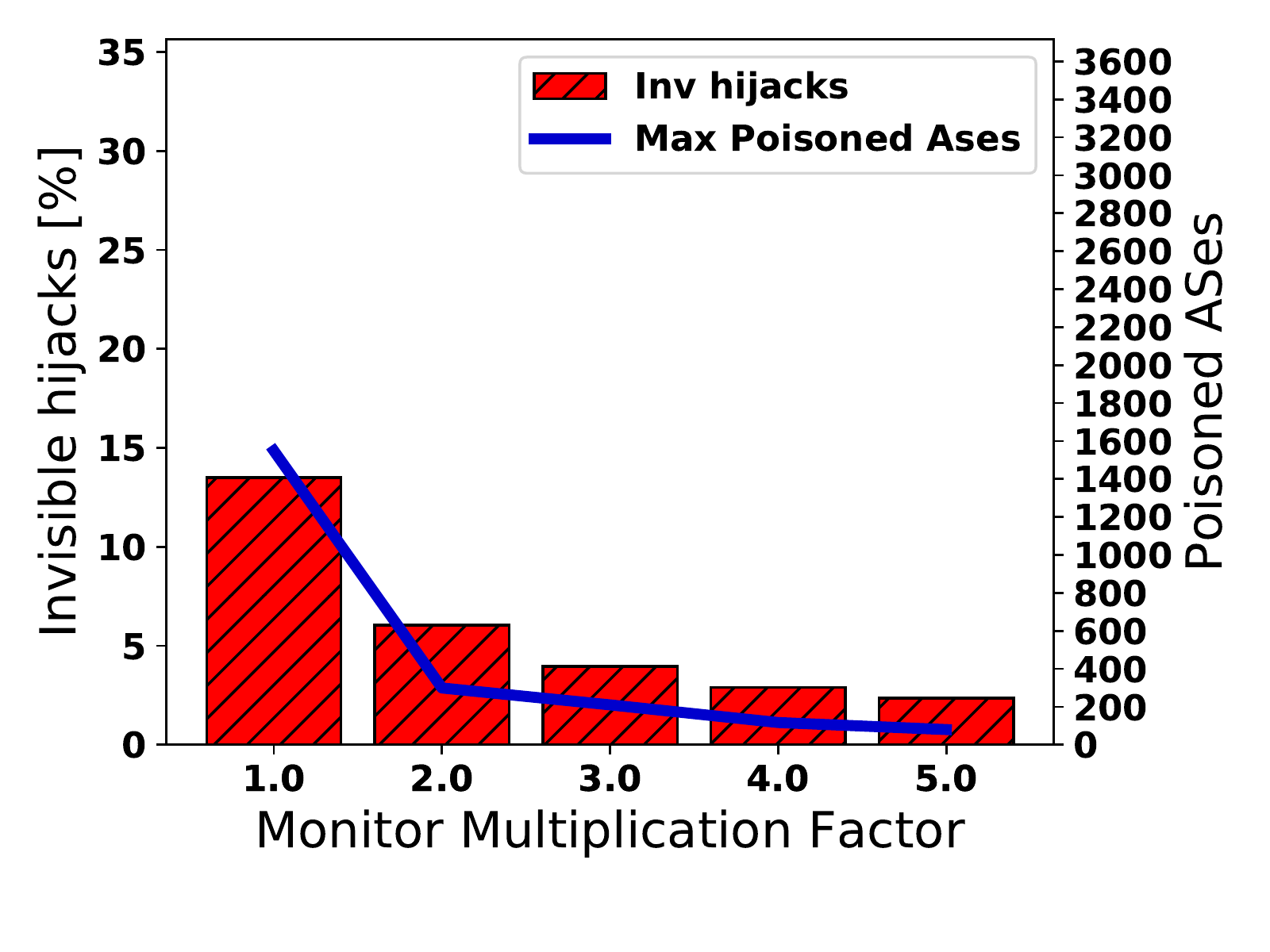}%
		\caption{Normal hijacker}
		\label{subfig:normal_hijack_bar}
	\end{subfigure}%
	\begin{subfigure}[t]{0.5\linewidth}
		\vskip 0pt 
		\includegraphics[width=\linewidth]{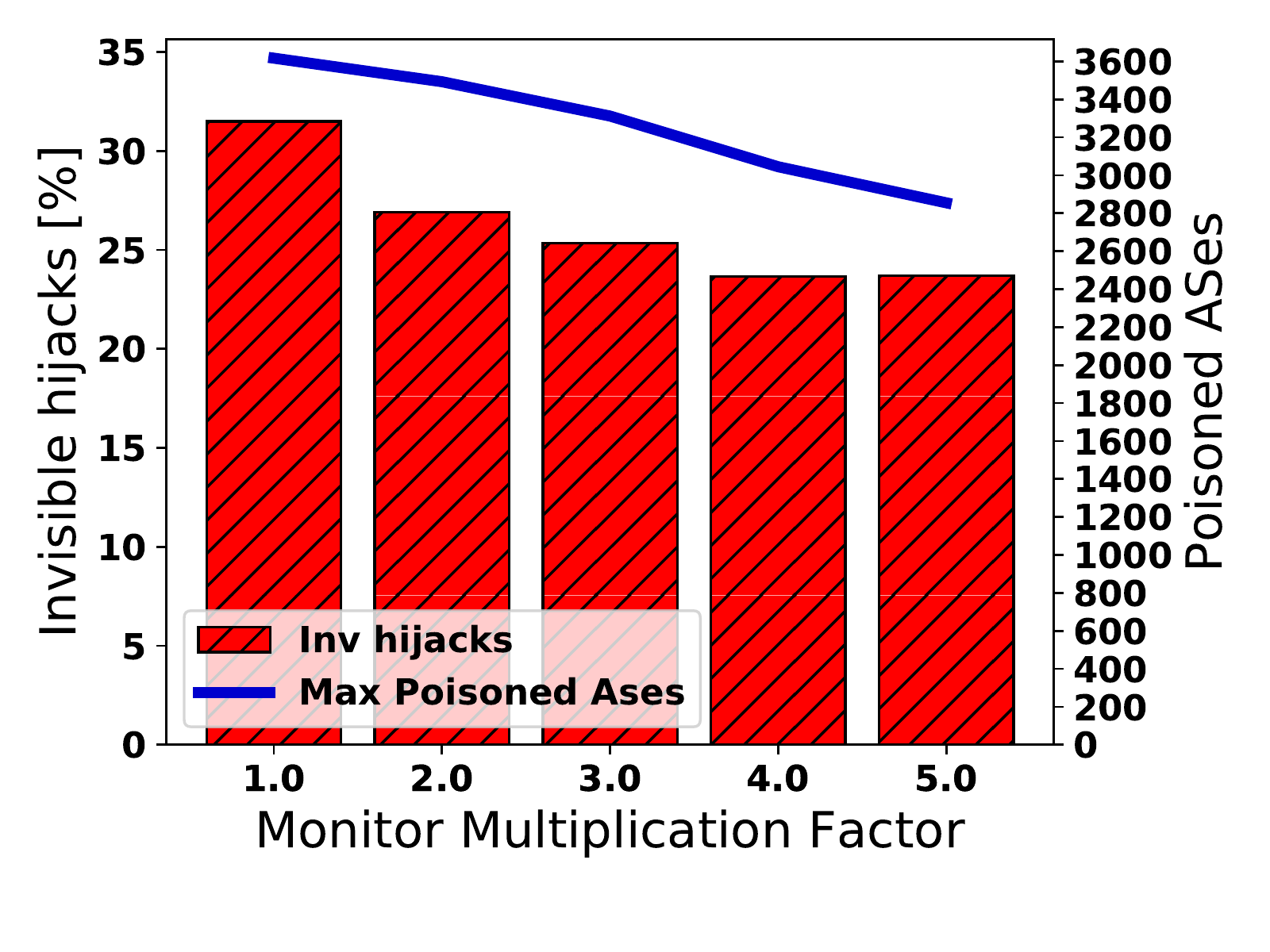}%
		\caption{Smart hijacker}
		\label{subfig:smart_hijack_bar}
	\end{subfigure}%
	\caption{Stealthy hijacks produced by a naive (baseline) hijacker versus our
	stealthy hijacker as we scale the number of existing monitors. The bars
	depict the number of stealthy hijacks that succeeded (out of 2000
	simulations), while the line shows the maximum number of poisoned/affected
	ASes for the corresponding scale of monitors.}%
	\label{fig:scale}%
\end{figure}

\myitem{Stealthy hijacks can often evade detection.} We compare the previously
introduced stealthy hijacker (called \textit{smart}) with a simpler one (called
\textit{normal}), which announces the victim prefix in an RPKI-valid manner. To
do so, we run 2000 simulations in which we pick a different victim and hijack AS
and verify whether the hijacker can attract traffic from other networks without
being detected by any of the $320$ BGP monitors used by Artemis.
\cref{subfig:normal_hijack_bar} on the left y-axis shows the percentage of
successful attacks for the normal hijacker undetected by the BGP monitors
(depending on how many are deployed). With a factor of one, \ie $320$ monitors,
the percentage of successfully invisible attacks is $14$\%, decreasing to less
than $1$\% when we deploy five times more monitors. On the right y-axis, we show
the maximum amount of networks affected by a successful invisible hijack (in the
order of thousands with $320$ monitors). We also observe (not shown in the
graph) that in 30\% of the attacks, more than one hundred networks have been
hijacked. Even if we have five times the amount of monitors, $5$ percent of the
attacks affect more than $300$ networks. \cref{subfig:smart_hijack_bar} shows
the same analysis for the stealthy hijacker (smart). In this case, the amount of
successfully invisible interception hijacks grows to $31$\%, and the maximum
amount of hijacked networks remains in the order of thousands even when the
number of deployed monitors increases five times. These results indicate that
BGP-based monitoring systems are insufficient to detect some more sophisticated
hijack types.

\section{Overview}
\label{sec:overview}

This section introduces \system's main components and characterizes the type of
hijack attacks it can detect.

\subsection{\system's main components}
\label{sec:componenets}

\begin{figure}
	\centering
	\includegraphics[width=1\columnwidth]{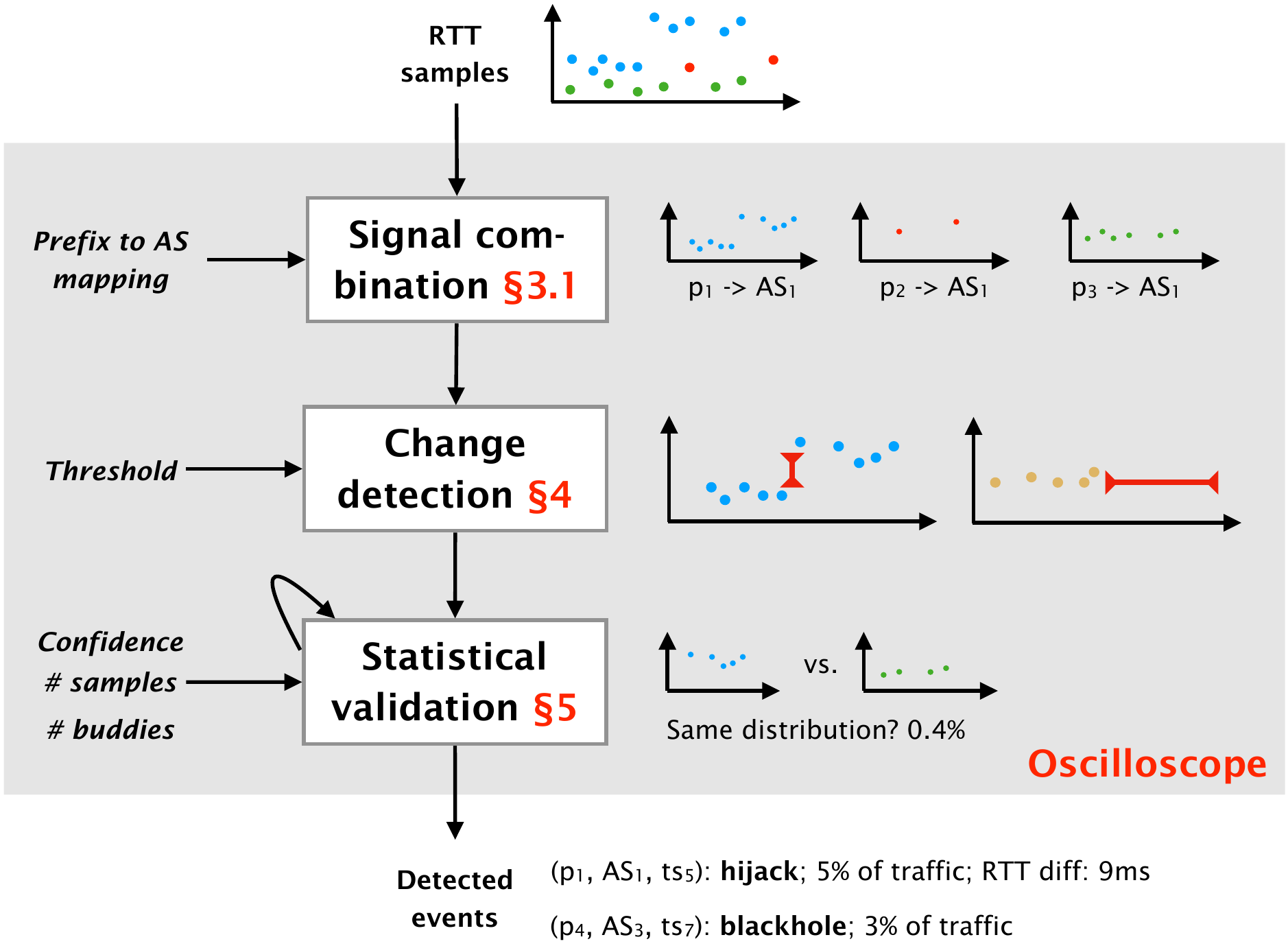}
    \caption{\system contains three major functional blocks. It combines RTT
    samples, then detects changes in the combined signals and performs
    statistical tests to validate the detected events.}
	\label{fig:pipeline}
\end{figure}

The \system pipeline comprises three main functional blocks
(\cref{fig:pipeline}): the combination of RTT signals, the detection of changes
in the signals, and the hijack validation.

\myitem{Input: RTT sample extraction.} The main input to \system is a set of RTT
samples extracted from the flows that traverse the network \system operates in.
There exist multiple methods to extract RTT signals from network traffic. For
example, Google's Espresso~\cite{espresso} provides such RTT samples from their
servers, and recent work has shown how to extract RTT samples from programmable
data-plane devices~\cite{rexford_rtt,dart}. Therefore, this work assumes that
the RTT signals have already been extracted.

Extracting and gathering RTT signals from all the flows in a large network may
become a scalability problem. In such cases, we can feed the RTT signals of only
the largest flows to \system. As we show later, \system does not need to monitor
all the flows from/to an IP prefix to detect a hijack.

\myitem{Stage 1: Signal combination.}
The first step of \system is to build a ``combined'' RTT signal that captures
changes at the level of entire IP prefixes and\slash or ASes. This level of
granularity allows us to detect a hijack attack that affects even a single
external AS, provided that we have ongoing traffic flows from\slash to that AS
network. The challenge is to combine flows \textit{(i)}\,efficiently while
\textit{(ii)}\,preserving the necessary information to detect potential hijacks.

\system takes as input a given set of internal IP prefixes to be monitored and
groups the RTT samples at the granularity of /24 IP prefixes. For instance, a
/23 IP prefix will result in two combined RTT signals corresponding to the two
/24 subnets, allowing us to detect both same-prefix and more-specific attacks.
In fact, common BGP filters~\cite{min_24} would block hijacks of a /25 IP prefix
or more specific ones. We note that aggregating the RTT signals of all the flows
from\slash to a monitored /24 is, however, still too coarse-grained, as hijacks
invisible to BGP monitors only affect certain Internet regions. Thus, we combine
RTT signals per \textit{i)}\,/24 prefix and \textit{ii)}\,per external AS with
which a flow communicates. In other words, each \emph{combined signal} contains
RTT samples from traffic exchanged between one of our /24 IP prefixes and all
the prefixes belonging to a \emph{single} external AS.

To figure out which IP prefixes belong to which AS, \system relies on the AS
path of the BGP routes received at the local routers. For each IP prefix, the
corresponding AS path usually indicates which AS originated the route

\myitem{Stage 2: Change detection.}
In the second block, \system analyzes each combined signal to detect sudden RTT
changes (increase or decrease). We present a simple overview and refer the
reader to \cref{sec:detection} for more details. We perform the change detection
by comparing long- and short-term minimum RTT values. Suppose the difference
exceeds a user-defined threshold (input parameter) and is ongoing for the entire
duration of a user-defined time window. In this case, \system detects an event
and triggers the last operation of the pipeline: It validates whether this event
is a BGP hijack or a genuine change in the Internet state. In addition, we also
forward combined signals to the validation stage if they suddenly terminate,
which could indicate a blackhole attack.

\myitem{Stage 3: Hijack validation.}
Several Internet routing events may lead to changes in the RTT of specific
flows, \eg reconfiguration, policy updates, load balancing, or fiber cuts. The
last block in \system's pipeline is therefore responsible for distinguishing
``true'' hijacks from normal routing events.

\pagebreak

\system validates hijacks using the concept of ``buddy'' prefixes which
translate insights from~\cite{buddyguard} into the data plane. We say that a set
of IP prefixes originated by the same AS are \textit{buddies} if their flows
towards the same external AS destination follow the same path. In the absence of
malicious hijack attacks, IP prefixes originated by the same AS propagate
through BGP along the same paths, as most ASes do not apply per-prefix
policies~\cite{gill2014survey}. Hence, buddies offer comparison points that can
identify hijacks: a hijack of one of our prefixes does not affect its buddies,
whereas normal routing events affect \emph{all} buddy prefixes similarly as they
traverse the same AS path. When the change detection block reports an event for
a particular IP prefix, \system collects RTT samples from the prefix and one of
its buddies. Using a two-sample statistical test, \system checks whether the two
RTT samples appear to come from the same \textit{distribution}; if not, we take
this as evidence that the flows started to follow different paths, which
indicates a potential hijack. If possible, \system repeats this validation step
with additional buddy prefixes to increase its confidence.

\subsection{Detectable hijacks}
\label{sec:capabilities}

\system detects hijacks based on changes in the observed data-plane signals,
which reveal blackhole and interception BGP hijacks. In a blackhole attack,
traffic from one or multiple external ASes targeted to some of our internal
prefixes is terminated/reaches a blackhole in an unwanted AS. An interception
attack is more refined. Instead of terminating the ongoing flows, an attacker
intercepts the traffic and forwards packets over different/additional ASes
before they reach their original destination, \ie our AS. Both hijack types emit
data-plane signals, which we can detect by analyzing RTT samples collected in
our AS. A blockhole hijack leads to a (one-directional) traffic loss, resulting
in an abrupt termination of any observable RTTs. In contrast, an interception
hijack leads to a potential change in the RTTs. Given \system's combined signals
that focus on a /24 level, Oscilloscope can use its detection pipeline for
same-prefix and more-specific hijack attacks. In summary, \system can detect all
hijack types introduced in our attacker model (\cref{sec:background}) but might
miss hijacks that do not generate a representative data-plane signal (no flows
or small RTT change), as we show in \cref{sec:evaluation}.

\section{Large-Scale Data-Plane Change Detection}
\label{sec:detection}

This section explains how \system detects changes in the combined RTT signals,
filters noise, and reacts to events where no RTT samples are available.

\subsection{Change detection}
\label{sec:det_cpd}

\begin{figure}
	\centering
	\includegraphics[width=1\columnwidth]{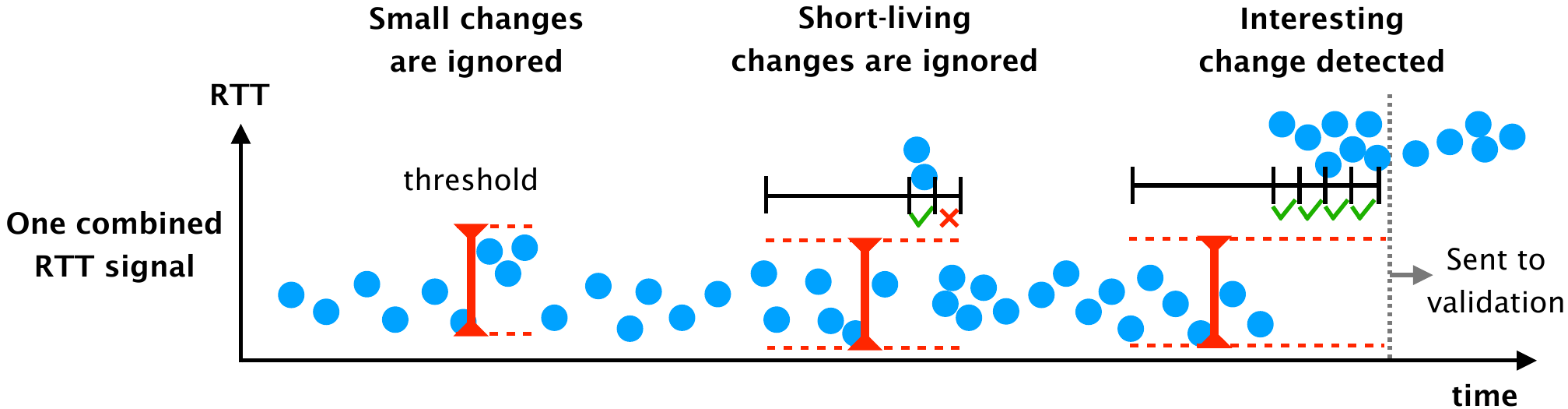}
	\caption{\system uses long- and short-term windows to detect
	minimum RTT changes larger than a given threshold.}
	\label{fig:change_workflow}
\end{figure}

\cref{fig:change_workflow} shows \system's workflow to detect changes in a
combined RTT signal using simplified examples. As explained in
\cref{sec:componenets}, a ``combined signal'' contains all RTT traffic samples
exchanged between one of our /24 prefixes and all the prefixes belonging to a
single external AS. The change detection process runs on each combined signal
separately once every second. That also means that the number of RTT samples
considered each second depends on the amount of traffic in a specific combined
signal.

\myitem{Using a threshold to detect large enough RTT changes.} RTT signals are
often noisy due to network congestion and end-host processing delays. This noise
may even be exacerbated when combining RTT samples from different flows, which
is the case when \system builds combined signals. Therefore, \system uses a
threshold (indicated in \cref{fig:change_workflow} with the red vertical
bars) to dismiss changes in the RTT signal that are too small (example on the
left). The network operator defines the threshold directly impacting the number
of detectable hijack events (as empirically validated in
\cref{sec:evaluation}).

\myitem{Using the minimum RTT over time windows to track the RTT.} However, we
do not apply the threshold for every two consecutive RTT samples but rather use
it to detect changes in the \emph{minimum} RTT observed in a long-term and
short-term window. \system uses the minimum RTT for two reasons. First, the
minimum is practical when considering RTT measurements, as there is an explicit
lower bound on how fast packets can traverse over a given path~\cite{aws-te}. As
such, a change in the observed long-term minimum RTT indicates a potential path
change. Second, \system can update the current minimum value whenever it
receives a new RTT sample and does not need to store all the samples belonging
to one combined signal. That improves efficiency and saves storage resources
which is essential as the number of combined signals is large depending on the
network size.

\myitem{Example.} \cref{fig:change_workflow} shows how \system uses short- and
long-term windows. In the example in the middle, \system realizes that the
short- and long-term RTT difference exceeds the given threshold for a few RTT
samples (indicated with the green tick below the short-term window). If \system
would now immediately report the change and continue with the validation step,
we would be very susceptible to short-living RTT spikes (\eg due to queues
filling up). For this reason, \system saves the long-term RTT minimum and
repeats the comparison in the next short-term windows. In the example, the RTT
difference no longer exceeds the threshold (red cross), and we dismiss the
change. \system only deems a change as significant if the RTT difference exceeds
the threshold in as many consecutive short-term windows as the long-term window
is wide. In our implementation, we use a four-second long-term window and a
short-term window of one second. For that reason, \system needs to detect the
change in four consecutive short-term windows before acknowledging the RTT
change (as shown in the example on the right in \cref{fig:change_workflow}). We
found out that a long-term window of four seconds is long enough to filter out
short delay spikes (due to congestion) while being short enough to allow \system
to operate reactively and detect potential hijacks in a timely manner.

\myitem{Detecting positive and negative RTT changes.} To detect potential hijack
events that would decrease the observed minimum RTT (i.e., a hijacker that can
provide better connectivity than the original path), \system compares the
absolute value of the minimum difference with the threshold. Also, it checks
that the sign of the difference is consistent in the four consecutive short-term
windows. If all these conditions hold, \system continues with the validation
step.

\subsection{Dealing with a lack of RTT samples}
\label{sec:no_samples}

It is important to note that \system handles short-term windows without any RTT
samples in a unique way. We observe short-term windows without samples if, for
example, the combined signal consists of flows with low activity or an event
that causes packet loss (which results in the loss of RTT samples). In such a
case \system performs two operations. First, it extends the long-term window by
one unit (\ie keeps track of the existing long-term RTT minimum) rather than
decreasing the amount of RTT samples in the long-term window with every empty
short-term one. Without that, \system would eventually lose the long-term
minimum RTT value, which an attacker could abuse to evade detection. More
precisely, an attacker could drop the initial hijacked packets before starting
an interception attack, and \system would no longer be able to detect the
introduced RTT change. Second, \system starts a blackhole detection process.
\system once again uses a similar approach to track the minimum RTT, but this
time it searches for four consecutive short-term windows \emph{without} any RTT
samples. However, to accommodate combined signals containing flows with
infrequent RTT samples (\ie some short-term windows without any samples), the
number of required consecutive short-term windows without any samples (4) is
multiplied by $n+1$ where $n$ corresponds to the number of empty short-term
windows in the currently considered long-term window. With this design choice,
we combine the well-working long-term window size of four with an adaptive
component (multiplication by $n+1$) based on the current behavior of the
combined signal. If enough consecutive short-term windows contain no samples,
\system forwards the potential blackhole to the validation stage.

\begin{figure}
	\centering
	\includegraphics[width=1\columnwidth]{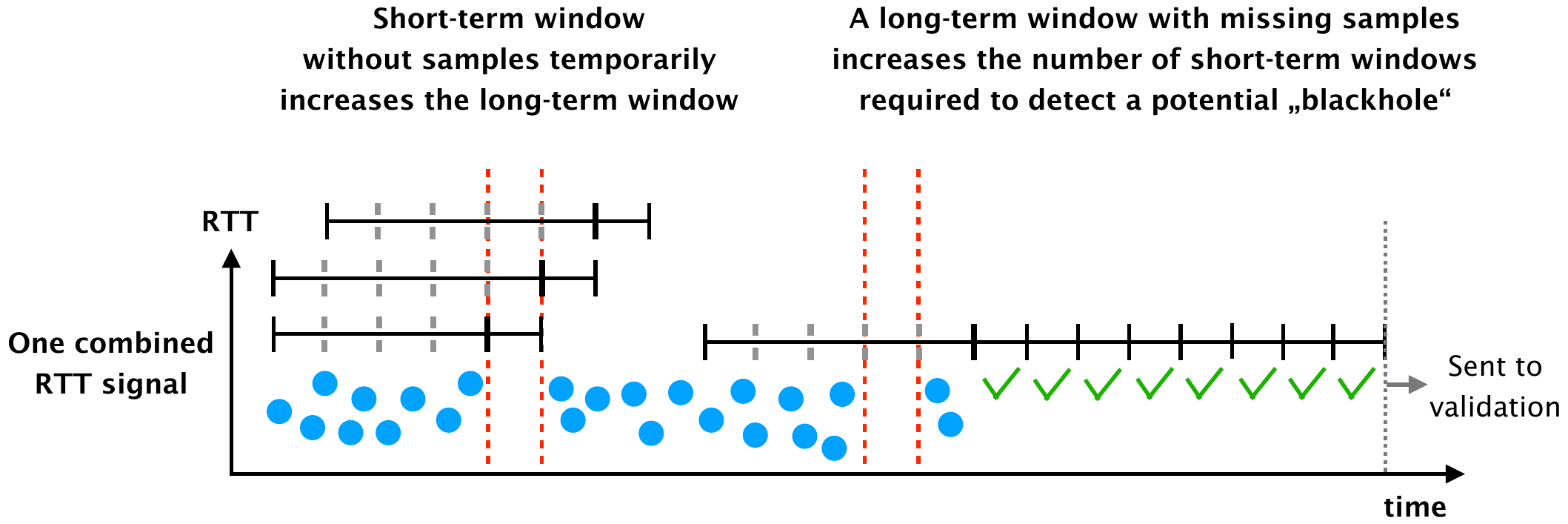}
	\caption{Short-term windows without any RTT samples increase the
	corresponding long-term windows and influence the detection process of
	potential blackholes.}
	\label{fig:blackhole_workflow}
\end{figure}

\myitem{Example.} \cref{fig:blackhole_workflow} illustrates these two
operations. In the example on the left, the current long-term window now
contains five instead of four short-term windows, as one does not contain any
RTT samples. The same holds for the example on the right, where \system requires
$4 \times (1+1) = 8$ consecutive short-term windows before the potential
blackhole observation is forwarded to the validation stage.

\section{Validation and Analysis of Changes}
\label{sec:validation}

We now introduce the concept of buddy prefixes and explain how we use
statistical tests to validate our hijack assumptions.

\subsection{Buddy concept}
\label{sec:buddies}

\system's primary validation strategy is built around the intuition that
successful interception (or blackhole) hijacks will introduce a forwarding
change reflected in the collected RTT signals. To have a baseline reference for
detecting hijack attacks, \system uses the concept of so-called ``buddy''
prefixes. In the absence of any hijacks, traffic from two buddy prefixes should
follow the same paths when communicating with identical external ASes. The
concept of ``buddy'' prefixes is not new. Buddyguard~\cite{buddyguard} first
introduced the term to describe prefixes that behave similarly (in the absence
of hijacks) when looking at them in the \textit{control plane}. \system
translates this concept to the \textit{data plane}.

To build buddy prefixes, we split the IP space advertised by our own AS into /24
prefixes, the smallest prefix size advertised by BGP
Internet-wide~\cite{min_24}. As a result, a potential hijacker must create fake
advertisements for at least one of these /24 prefixes to influence our traffic.
Consequently, the hijack will equally affect traffic inside one buddy prefix. We
leave the problem of identifying buddy prefixes in real-time as future work. Our
analysis with real BGP data in \cref{sec:deployment} shows that most IP prefixes
have multiple buddies.

\begin{figure}
	\centering
	\includegraphics[width=0.9\columnwidth]{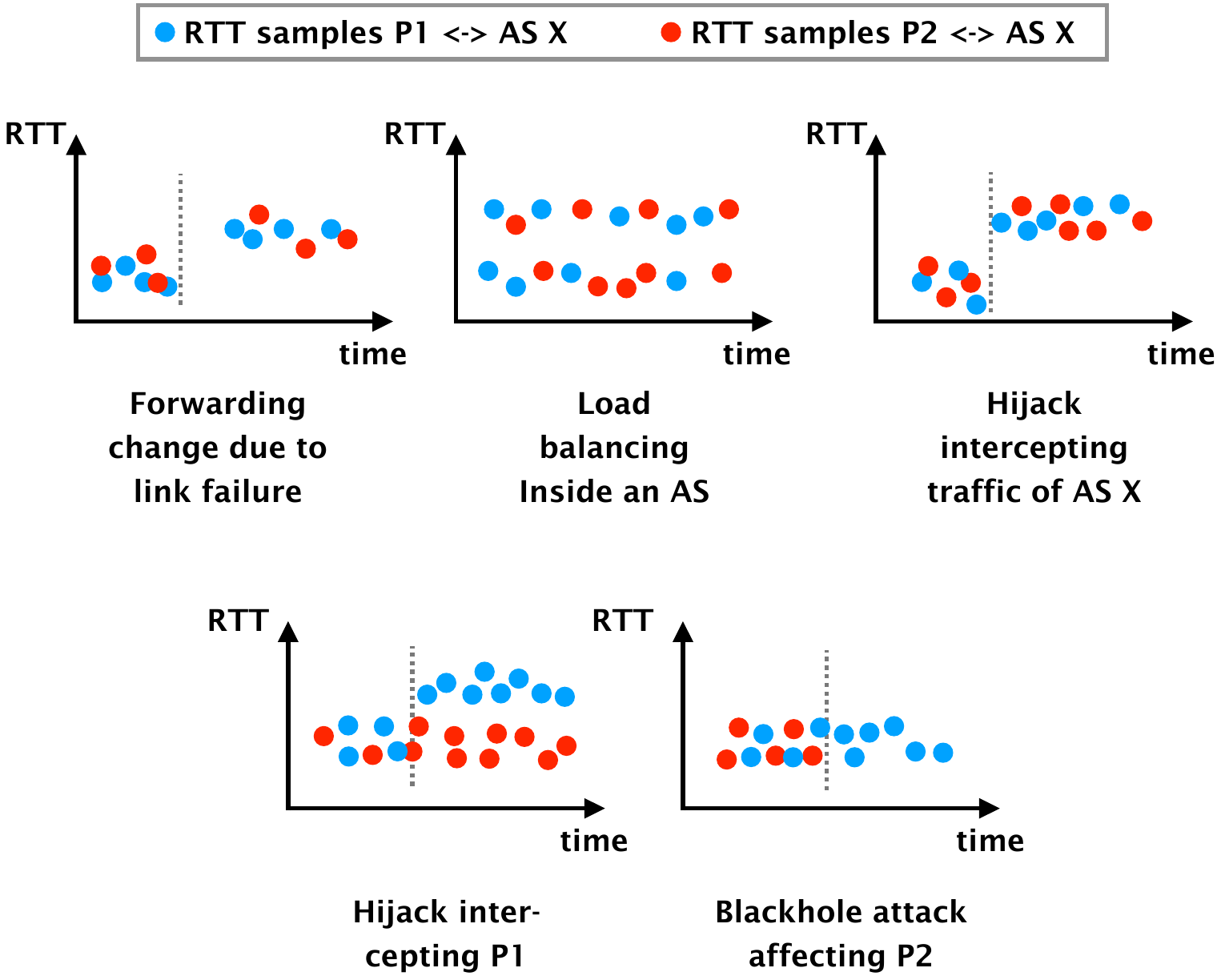}
    \caption{RTT samples of two buddy prefixes (P1, P2) exchanging traffic
    with AS X. We only observe an asymmetric behavior between buddy samples if a hijack affects P1 or P2.}
	\label{fig:buddy_example}
\end{figure}

\cref{fig:buddy_example} shows simplified RTT distributions for various
network events. Most of these examples are based on traces used in our
evaluation (\cref{sec:exp}). Buddy prefixes (P1 and P2) exchange traffic with
AS X. The first two examples (forwarding change due to link failure and load
balancing) show that RTT samples from both buddies are equally affected. That
means, eventually, flows from both buddies will use the new forwarding path
after the link failure or use the same set of physical paths when load balanced,
respectively. The same observation holds in the third case (an interception
hijack targeting prefixes of AS X), where traffic from P1 \emph{and} P2 will be
affected. Note that \system does \emph{not} try to detect such hijacks as the
hijack does not target one of \emph{our} prefixes. Finally, the interception
attack of P1 and the blackhole attack targeting P2 (lower two examples) are the
only two events leading to observable asymmetries in the RTT distributions of
the two buddy prefixes. \system is precisely looking for such changes and
compares samples between multiple buddy prefixes to validate its hijack
assumptions in these cases.

\subsection{Sample collection}
\label{sec:sample_collection}

As soon as \system detects a change in traffic belonging to a combined signal
(\cref{sec:detection}), it collects RTT samples from this combined signal. In
addition, we also look for comparison samples from valid buddy prefixes.
Initially, \system assumes that all our /24 prefixes are buddies (compare
\cref{sec:deployment} and \cref{sec:discussion}). However, \system cannot
blindly collect comparison samples from these buddy prefixes. The problem is
that \system neither knows a priori how many /24 prefixes are affected by the
potential hijack (\ie a hijack targeting a /23 prefix affects two /24 buddies)
nor do we want to make any assumptions. Suppose \system would pick validation
samples from a buddy also targeted by the hijack. In that case, the validation
step might report wrong results because we do not have a valid comparison point.

For this reason, \system \emph{ignores} buddies that \textit{(i)}\,were recently
declared as hijacked towards the currently observed external AS (an operator can
clear these buddies after investigating the event); or \textit{(ii)}\,are
currently also in the process of being validated as we detected a change in
their corresponding combined signal towards the same external AS (remember that
we perform the change detection for each combined signal individually). The
second criterion includes buddies for which the change is \emph{nearly}
confirmed, \ie we already found the change in three of the required four
consecutive short-term windows (see \cref{sec:detection}), to prevent strict
time synchronization requirements between the change detection processes. From
the remaining buddy prefixes, \system follows a ``first come, first served''
principle to pick the required amount of comparison samples from the specified
number of buddy prefixes (two operator-defined parameters).

Note that all buddy prefixes are ignored for apparent events (\eg a forwarding
change introducing a significant RTT change). In this case, \system correctly
terminates the validation process and does not report any hijacks. In the more
likely cases (\eg changes detected due to noise, hijacks targeting specific
prefixes, \dots) \system applies the statistical test described in the following
subsection. It compares the distributions of samples collected from the combined
signal that observed the change and its buddy prefixes.

If \system tries to validate a potential blackhole, it expects to find no
samples from the corresponding combined signal. We will immediately discard the
blackhole assumption if we still observe a sample. Suppose, in addition, the
buddy prefix(es) also do not provide any RTT samples (\ie no more traffic
towards this external AS), and we cannot confirm the blackhole idea. It looks
like a total traffic loss, \eg due to a link failure, and \system will not
report anything.

\subsection{Statistical tests}
\label{sec:test}

After collecting samples from the combined signal, which observes a change and
at least one buddy prefix, \system uses the Wilcoxon-Mann-Whitney test
(e.g.,~\cite[p.~128+]{stats_book}) to compare the two sample sets. The
Wilcoxon-Mann-Whitney test is a non-parametric statistical test that only
assumes that the samples from both sets are independent (given as they come from
different /24 prefixes) and that we can order them (\ie one RTT sample is bigger
than another one). We selected this test for at least two reasons. First, given
that it belongs to the class of non-parametric tests, we do not require that the
collected RTT samples follow additional assumptions, \eg that they belong to a
specific distribution. Second, we can also apply the test if the number of found
RTT samples is small (compare~\cite[Appendix Table J]{stats_book}), which is
often problematic for comparable parametric tests.

As null hypothesis $H_0$, we assume that the two sample-set distributions are
equal. In contrast, the alternative hypothesis $H_1$ states that one sample
distribution is stochastically larger (or smaller) than the other. An
interception hijack will most likely increase the observed RTT but could also
decrease it. For this reason, we perform a \emph{two-sided} test.

The test first sorts all collected RTT samples in increasing order and then sums
up the ranks of samples belonging to the same set (\ie ranks of all samples
coming from the buddy prefix). If both sample sets belong to the same
distribution ($H_0$), the rank sum should roughly be equal.

The network operator defines a confidence level as an input parameter to
\system's validation step. Whenever the test result is above the confidence
level, we assume that the null hypothesis holds and the two sample sets come
from the same distribution, \ie they follow the same forwarding path. Otherwise,
we assume they come from two different distributions, strengthening our
confidence that we observe an ongoing hijack. If available, \system repeats the
same test with samples from additional buddy prefixes.

If the validation test indicates that the null hypothesis holds, \system
immediately discards the event and will not report anything. However, if the
tests suggest that the RTT samples come from different distributions, we can
optionally repeat the validation step once (or multiple times). That means
\system will collect another batch of RTT samples from the combined signal
observing the change and its buddy prefixes. Not only will that strengthen our
hijack assumption, but it also allows us to compare against other buddy
prefixes, diversifying the collected samples. As we show in
\cref{sec:accuracy}, repeating the validation step once greatly reduces
the number of false positives/increases \system's precision.


\section{Experimental Setup}
\label{sec:exp}

We face two challenges when evaluating the hijack-detection performance of
\system. First, we require a realistic Internet-like virtual environment, which
is notoriously hard to obtain. And second, we need an extensive dataset of
hijack events (with corresponding data-plane information) which, to the best of
our knowledge, does not exist. Testbeds such as the PEERING~\cite{peering}
platform have only limited prefixes and ASes under their control. This section
describes how we address these two challenges.

\subsection{Mini-Internet emulation}
\label{subsec:emulation}

To evaluate \system, we need an environment emulating
\begin{itemize}[nosep,leftmargin=*]
	\item Multiple ASes with BGP connectivity;
	\item Realistic BGP relationships between ASes;
	\item Inter-domain links with realistic delays;
	\item Prefix advertisements and traffic generation from end hosts.
\end{itemize}

\noindent We use the mini-Internet platform~\cite{mini-internet} as a basis for
our emulation. The mini-Internet runs each device (\eg router, host) in its own
Docker container and links them together to enable Internet-like connectivity.
We build an AS topology extracted from the real Internet using the RIPE Atlas
platform~\cite{ripe2020ripe}. We collect the built-in IPv4 traceroute
measurements from a global network of probes, each of which automatically
performs those measurements at 20-30 minute intervals towards the 13 DNS root
servers. Using the IPv4 Paris-enabled traces, which are protected against common
load balancing problems~\cite{augustin2006avoiding}, we extract the inter- and
intra-domain links observed by each probe. We also estimate the link delay
between two destinations, $A$ and $B$, as $delay_{A \rightarrow B} = RTT_B -
RTT_A$.

One day of measurements from RIPE contains about $149$M traceroutes involving
$520$k intra-domain and $116$k inter-domain links interconnecting $12$k ASes.
The mini-Internet platform does not scale to networks that large (when running
on a single server). Thus we keep only the most frequently discovered
inter-domain links in our topology. Keeping only the 3\% most commonly found
links yield $2303$ links interconnecting $968$ ASes, out of which the most
significant connected subset contains more than $600$ ASes, which we use for our
emulated topology. Finally, we use the AS-relation
dataset~\cite{CaidaASrelations} to obtain the BGP relationships between the ASes
in our topology. The internal AS topology is abstracted as one router connected
to one or multiple hosts, which act as endpoints for the advertised prefixes.
The ASes build eBGP connections with their neighbors following the extracted
relationships. For all the external links between the ASes we use
\texttt{tc-netem}~\cite{tc_netem} to apply link delays and jitters based on the
collected traceroute measurements. If we do not have enough data to estimate a
link delay, we default to a delay of 5\,ms and 2\,ms jitter.

These extracted links, delays, and relationships are configured in the
mini-Internet to provide a realistic Internet emulation environment for our
evaluation.

\subsection{Hijack dataset}
\label{subsec:dataset}

We leverage our emulated mini-Internet~(\cref{subsec:emulation}) to generate a
synthetic dataset of packet traces containing hijacks. First, we identify a set
of possible hijack events for our topology. Then we collect packet traces for
simulated traffic from multiple runs with and without hijack events.

First, we select one well-interconnected AS as ``our AS''; it advertises twelve
different /24 prefixes in the entire mini-Internet. We then choose 70 other ASes
to perform an interception attack towards our prefix space. For each of these
ASes, we generate one hijack event where the AS hijacks one or two of our /24
prefixes from one or two of its directly connected neighbors. The hijacking AS
uses path poisoning~\cite{poisoning}, and thus, the fake advertisement can
propagate in a localized part of the mini-Internet and potentially hijack
traffic from additional ASes. Similarly, we generate more than a hundred
blackhole events by simply dropping the attracted traffic rather than forwarding
it back to our AS. Finally, we also generate ``normal'' forwarding events, such
as link failures, by dropping all the traffic on a given link. The routers run
Bidirectional Forwarding Detection (BFD)~\cite{bfd} to detect link failures and
trigger corresponding actions. As a result, such a link failure either leads to
a complete loss of traffic or a benign forwarding change with corresponding
increases or decreases in the observed RTT. For the evaluation, knowing how a
specific event influences the observable data-plane signals (i.e., ground-truth
behavior) is essential. Therefore, we compare traceroute outputs before and
after each event to extract all pairs of /24 prefixes and external ASes that
observe a forwarding path change due to the event. Additionally, we can quantify
the induced RTT change. Obviously, \system does not have access to this
information. We only used it to label and group the events for the following
evaluation.

We can now collect packet traces containing these hijacks or normal network
events. We use Flowgrind~\cite{flowgrind} to generate traffic mimicking real
applications, \eg video streaming or HTTP-style request-response patterns. We
generate an average of 100 flows resulting in about 200MBps or 70k packets per
second from our /24 prefixes towards ASes affected by the various events. We use
tcpdump~\cite{tcpdump} to record the packets observed on all interfaces
connected to the hosts in our AS. Each experiment runs for two minutes, with the
event starting after one minute and staying active until the end of the run. We
perform one run for each event in our dataset and 30 runs without any specific
events, which results in more than six hours of traces. For each trace, we
extract the RTTs of all TCP flows by matching sequence and acknowledgment
(SEQ/ACK) numbers (similar to the method in~\cite{rexford_rtt}).

Equipped with this set of RTTs traces, we can now evaluate the hijack-detection
performance of \system \emph{offline} for different choices of parameters
(\cref{sec:evaluation}).

\section{Evaluation}
\label{sec:evaluation}
This section first evaluates \system's accuracy looking at precision and recall
values for different input parameters. With a detection threshold of 3\,ms
\system reaches a 94\% recall and 93\% precision value when analyzing all events
in our generated data set containing more than six hours of traces. Afterwards,
we discuss \system's detection speed and perform some micro benchmarks. Finally,
we evaluate our buddy concept with control-plane data from the real Internet.

\subsection{\system's accuracy}
\label{sec:accuracy}

The core function of \system is its ability to detect hijacks and differentiate
them from normal network events. Performance is mainly affected by two system
parameters: the detection threshold (\cref{sec:det_cpd}) and the number of
repetitions of the validation step (\cref{sec:test}). In the following
experiments, we quantify how these affect \system's precision (\ie the rate of
true positives among the detected events) and recall (\ie the rate of true
events correctly detected).

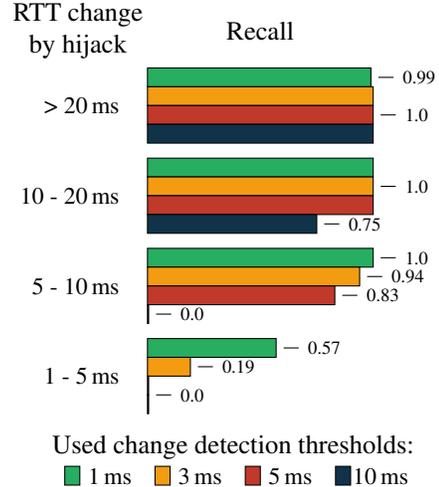
\begin{figure}
\centering

\begin{tikzpicture}[every node/.style={scale=1}]

    \newcommand\SizeFull{3}

    \newcommand\BarHeight{0.25cm}
    \newcommand\RowHeight{1.2cm}
    \newcommand\LabelGap{-0.25}

    \newcommand\LabelGapBar{0.1}

    \newcommand\FirstFull{1*\SizeFull}
    \newcommand\FirstLess{0.99*\SizeFull}

    \newcommand\SecondFull{1*\SizeFull}
    \newcommand\SecondLess{0.75*\SizeFull}

    \newcommand\ThirdFull{1*\SizeFull}
    \newcommand\ThirdMiddle{0.94*\SizeFull}
    \newcommand\ThirdLess{0.83*\SizeFull}
    \newcommand\ThirdZero{0.005*\SizeFull}

    \newcommand\FourthMiddle{0.57*\SizeFull}
    \newcommand\FourthLess{0.19*\SizeFull}
    \newcommand\FourthZero{0.005*\SizeFull}

    \newcommand\FirstFullLabel{1.0}
    \newcommand\FirstLessLabel{0.99}

    \newcommand\SecondFullLabel{1.0}
    \newcommand\SecondLessLabel{0.75}

    \newcommand\ThirdFullLabel{1.0}
    \newcommand\ThirdMiddleLabel{0.94}
    \newcommand\ThirdLessLabel{0.83}
    \newcommand\ThirdZeroLabel{0.0}

    \newcommand\FourthMiddleLabel{0.57}
    \newcommand\FourthLessLabel{0.19}
    \newcommand\FourthZeroLabel{0.0}

    \begin{scope}[]
        \begin{scope}[yshift=1cm]
            \node[align=center] at (3.7*\LabelGap , -2*\BarHeight) {RTT change \\by hijack};
            \node[align=center] at (0.5*\SizeFull , -2*\BarHeight) {Recall};
        \end{scope}

        \begin{scope}[yshift=0cm]
            \node[left] at (\LabelGap , -2*\BarHeight) {\small \textgreater\ 20\,ms};

            \draw[fill=cGreen] (0, 0) rectangle (\FirstLess, -\BarHeight);
            \draw[fill=cYellow] (0, -\BarHeight) rectangle (\SizeFull, -2*\BarHeight);
            \draw[fill=cRed] (0, -2*\BarHeight) rectangle (\SizeFull, -3*\BarHeight);
            \draw[fill=cDarkBlue] (0, -3*\BarHeight) rectangle (\SizeFull, -4*\BarHeight);

            \draw (\FirstLess+\LabelGapBar, -0.5*\BarHeight) -- (\FirstLess+\LabelGapBar, -0.5*\BarHeight) --++ (0.2, 0) node[right]{\scriptsize \FirstLessLabel};
            \draw (\FirstFull+\LabelGapBar, -2.5*\BarHeight) -- (\FirstFull+\LabelGapBar,  -2.5*\BarHeight) --++ (0.2, 0) node[right]{\scriptsize \FirstFullLabel};

        \end{scope}

        \begin{scope}[yshift=-1*\RowHeight]
            \node[left] at (\LabelGap , -2*\BarHeight) {\small 10\ -\ 20\,ms};

            \draw[fill=cGreen] (0, 0) rectangle (\SecondFull, -\BarHeight);
            \draw[fill=cYellow] (0, -\BarHeight) rectangle (\SecondFull, -2*\BarHeight);
            \draw[fill=cRed] (0, -2*\BarHeight) rectangle (\SecondFull, -3*\BarHeight);
            \draw[fill=cDarkBlue] (0, -3*\BarHeight) rectangle (\SecondLess, -4*\BarHeight);

            \draw (\SecondFull+\LabelGapBar, -1.5*\BarHeight) -- (\SecondFull+\LabelGapBar, -1.5*\BarHeight) --++ (0.2, 0) node[right]{\scriptsize \SecondFullLabel};
            \draw (\SecondLess+\LabelGapBar, -3.5*\BarHeight) -- (\SecondLess+\LabelGapBar,  -3.5*\BarHeight) --++ (0.2, 0) node[right]{\scriptsize \SecondLessLabel};

        \end{scope}

        \begin{scope}[yshift=-2*\RowHeight]
            \node[left] at (\LabelGap , -2*\BarHeight) {\small 5\ -\ 10\,ms};

            \draw[fill=cGreen] (0, 0) rectangle (\ThirdFull, -\BarHeight);
            \draw[fill=cYellow] (0, -\BarHeight) rectangle (\ThirdMiddle, -2*\BarHeight);
            \draw[fill=cRed] (0, -2*\BarHeight) rectangle (\ThirdLess, -3*\BarHeight);
            \draw[fill=cDarkBlue] (0, -3*\BarHeight) rectangle (\ThirdZero, -4*\BarHeight);

            \draw (\ThirdFull+\LabelGapBar, -0.5*\BarHeight) -- (\ThirdFull+\LabelGapBar, -0.5*\BarHeight) --++ (0.2, 0) node[right]{\scriptsize \ThirdFullLabel};
            \draw (\ThirdMiddle+\LabelGapBar, -1.5*\BarHeight) -- (\ThirdMiddle+\LabelGapBar, -1.5*\BarHeight) --++ (0.2, 0) node[right]{\scriptsize \ThirdMiddleLabel};
            \draw (\ThirdLess+\LabelGapBar, -2.5*\BarHeight) -- (\ThirdLess+\LabelGapBar, -2.5*\BarHeight) --++ (0.2, 0) node[right]{\scriptsize \ThirdLessLabel};
            \draw (\ThirdZero+\LabelGapBar, -3.5*\BarHeight) -- (\ThirdZero+\LabelGapBar,  -3.5*\BarHeight) --++ (0.2, 0) node[right]{\scriptsize \ThirdZeroLabel};

        \end{scope}

        \begin{scope}[yshift=-3*\RowHeight]
            \node[left] at (\LabelGap , -2*\BarHeight) {\small 1\ -\ 5\,ms};

            \draw[fill=cGreen] (0, 0) rectangle (\FourthMiddle, -\BarHeight);
            \draw[fill=cYellow] (0, -\BarHeight) rectangle (\FourthLess, -2*\BarHeight);
            \draw[fill=cRed] (0, -2*\BarHeight) rectangle (\FourthZero, -3*\BarHeight);
            \draw[fill=cDarkBlue] (0, -3*\BarHeight) rectangle (\FourthZero, -4*\BarHeight);

            \draw (\FourthMiddle+\LabelGapBar, -0.5*\BarHeight) -- (\FourthMiddle+\LabelGapBar, -0.5*\BarHeight) --++ (0.2, 0) node[right]{\scriptsize \FourthMiddleLabel};
            \draw (\FourthLess+\LabelGapBar, -1.5*\BarHeight) -- (\FourthLess+\LabelGapBar, -1.5*\BarHeight) --++ (0.2, 0) node[right]{\scriptsize \FourthLessLabel};
            \draw (\FourthZero+\LabelGapBar, -3*\BarHeight) -- (\FourthZero+\LabelGapBar,  -3*\BarHeight) --++ (0.2, 0) node[right]{\scriptsize \FourthZeroLabel};

        \end{scope}

        \begin{scope}[yshift=-4*\RowHeight]
            \node[align=center] at (0.38*\SizeFull , -1*\BarHeight) {Used change detection thresholds:};

            \draw[fill=cGreen] (-1.1, -2*\BarHeight) rectangle (-0.9, -3*\BarHeight);
            \node[align=left] at (-0.5 ,-2.5*\BarHeight) {\small 1\,ms};
            \draw[fill=cYellow] (-1.1+1.2, -2*\BarHeight) rectangle (-0.9+1.2, -3*\BarHeight);
            \node[align=left] at (-0.5+1.2 ,-2.5*\BarHeight) {\small 3\,ms};
            \draw[fill=cRed] (-1.1+2.4, -2*\BarHeight) rectangle (-0.9+2.4, -3*\BarHeight);
            \node[align=left] at (-0.5+2.4 ,-2.5*\BarHeight) {\small 5\,ms};
            \draw[fill=cDarkBlue] (-1.1+3.6, -2*\BarHeight) rectangle (-0.9+3.6, -3*\BarHeight);
            \node[align=left] at (-0.5+3.6 ,-2.5*\BarHeight) {\small 10\,ms};
        \end{scope}

    \end{scope}
\end{tikzpicture}

\caption{\system' recall for different change detection threshold values, binned
by the RTT change induced by the interception event. \system reliably detects
hijack events with RTT changes larger than the change detection threshold.}

\label{fig:detected}

\end{figure}

\subsubsection{Recall vs. magnitude of the RTT change}
\label{subsec:magnitude}

Setting the change detection threshold defines the detectable events (\ie
achievable recall). Intuitively, hijacks that induce RTT changes smaller than
the threshold are hard to detect.

\myitem{Setup.}
For each interception event in our dataset (\cref{subsec:dataset}), we run
\system with change detection threshold values of $\{1,3,5,10\}$ms and compute
the ratio of true events detected by \system (\ie the recall).

\myitem{Results.}
\cref{fig:detected} shows the recall for the different threshold values, binned
to the magnitude of the hijack-induced RTT change. As expected, the recall drops
when the RTT change gets close to the threshold. The detection becomes
unreliable for the smallest changes (1-5\,ms) (57\% detected with a threshold
value of 1\,ms). This is expected as the RTT change is smaller than the average
RTT signal noise, and \system cannot detect these hijacks.

Interestingly, in some cases, a threshold of 1\,ms fails to detect hijacks
inducing more than 20\,ms RTT changes. Closer inspection reveals that these
cases have only one buddy prefix available. At the same time, this buddy is
blacklisted for buddy comparisons due to a wrongly detected hijack event at a
previous point in the trace (compare~\cref{sec:sample_collection}). Thus when
the actual hijack occurs, \system does not find any buddy prefix for the
validation and hence does not report the hijack.

\myitem{Conclusion.}
\system reliably detects hijack events when they induce RTT changes larger than
the detection threshold. However, setting a threshold lower than the typical RTT
noise is not useful (it does not help to detect events inducing very low RTT
changes) and can even be harmful (\system fails to detect events that can be
caught with a larger threshold).

\subsubsection{Precision and recall vs. number of validations}
\label{subsec:num_validations}

Detecting RTT changes is relatively easy, but differentiating hijacks from
normal network events is more challenging. For this purpose, \system relies on
buddy prefixes to validate that an event is a true
hijack~(\cref{sec:validation}). One can perform the validation steps multiple
times; \system will report a hijack only if all validations agree that the event
is indeed a hijack. Thus, performing more validations will tend to decrease the
recall while improving the precision.

\begin{figure}
\centering

\begin{tikzpicture}[every node/.style={scale=0.85}]

  \newcommand\PlotWidth{240pt}
  \newcommand\PlotHorSep{46pt}
  \newcommand\PlotHeight{4cm}
  \newcommand\PlotVerSep{1cm}

  \begin{groupplot} [
    group style={
      group size=1 by 2,
      y descriptions at=edge left,
      horizontal sep=\PlotHorSep,
      vertical sep=\PlotVerSep,
      group name=weight group,
    },
    axis y line = left,
    ]


    \nextgroupplot[
      height=\PlotHeight,
      width=\PlotWidth,
      ylabel = {Recall},
      axis x line* = bottom,
      ymin = 0.84,
      ymax = 1.01,
      xtick = {0, 1, 2, 3, 4, 5, 6, 7, 8, 9},
      xticklabels = {10, 9, 8, 7, 6, 5, 4, 3, 2, 1},
      xmin = -0.2,
      xmax = 9.2,
      legend cell align=left,
      legend columns=3,
      legend style={
        draw=none,
        fill=none,
        font=\footnotesize,
        anchor=north,
        at={(0.6, 1.3)},
        /tikz/every even column/.append style={column sep=6pt},
      },
    ]

    \addplot+[mark=o, cRed, solid] table[x=x, y=y, col sep=comma] {tikz/recall_1.csv};
    \addlegendentry{one}

    \addplot+[mark=otimes, cYellow, solid] table[x=x, y=y, col sep=comma] {tikz/recall_2.csv};
    \addlegendentry{two}

    \addplot+[mark=square, cGreen, solid] table[x=x, y=y, col sep=comma] {tikz/recall_3.csv};
    \addlegendentry{three}


    \nextgroupplot[
        height=\PlotHeight,
        width=\PlotWidth,
        ylabel = {Precision},
        axis x line* = bottom,
        xlabel = {Threshold value [ms]},
        ymin = 0.7,
        ymax = 1.02,
        xtick = {0, 1, 2, 3, 4, 5, 6, 7, 8, 9},
        xticklabels = {10, 9, 8, 7, 6, 5, 4, 3, 2, 1},
        xmin = -0.2,
        xmax = 9.2,
      ]
  
      \addplot+[mark=o, cRed, solid] table[x=x, y=y, col sep=comma] {tikz/precision_1.csv};
  
      \addplot+[mark=otimes, cYellow, solid] table[x=x, y=y, col sep=comma] {tikz/precision_2.csv};

      \addplot+[mark=square, cGreen, solid] table[x=x, y=y, col sep=comma] {tikz/precision_3.csv};
  
  \end{groupplot}

  \draw (60pt, 93pt-9pt) node[left] {\footnotesize Number of validations:};

\end{tikzpicture}
\caption{Precision and recall values for different change detection thresholds
and different numbers of validation steps. Performing the validation twice
largely improves the precision without hurting the recall much. The third
validation has little effect. The best performance is obtained for a threshold
of 3\,ms, which is small but larger than the RTT signal noise.}

\label{fig:precision_recall}

\end{figure}
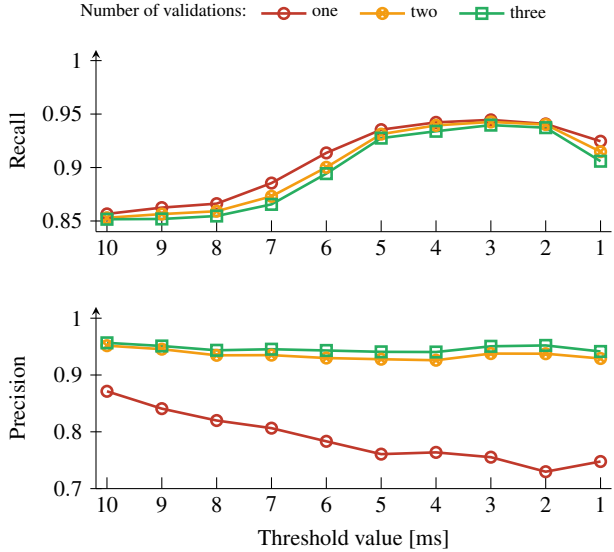

\myitem{Setup.}
We run \system for all events in our dataset with different change detection
threshold values~(1 to 10\,ms) and one, two, or three validation repetitions; we
compute the resulting precision and recall values.

\myitem{Results.}
\cref{fig:precision_recall} shows precision and recall values for the different
settings. Looking first at the precision, we observe a significant improvement
from one to two validations (up to 20\%), then a smaller gain with a third one.
This matches our expectations: since multiple validations must agree to identify
an event as a hijack, more validations reduce the number of false positives (\ie
improve precision). Conversely, the recall degrades with more validations. This
is also expected since true events have more chances of being wrongly discarded
as ``non-hijack.'' Yet, the recall loss is small (less than 2\%).

Let us now look at the performance when varying the detection threshold.
Generally, smaller threshold values detect more events that either result in
true or false positives during the validation step. As a result, smaller
thresholds degrade precision and improve recall. However, as discussed in
\cref{subsec:magnitude}, threshold values smaller than the noise level (1-2\,ms)
yield more false positives and limit the buddy prefixes available for
validation, thus hurting the recall as well.

One may notice a tipping point around 5\,ms. We explain this by the design of
our emulation setup, where we use a link delay of 5\,ms as a default
value~(\cref{subsec:emulation}). Our dataset is thus biased towards events
inducing a 5\,ms RTT change.

\myitem{Conclusion.}
Overall on our dataset, the best performance is obtained with a threshold of
3\,ms and two validations. This results in a 94\% recall and 93\% precision. A
third validation trades slightly better precision for slightly worse recall.

\subsubsection{Precision and recall vs. number of buddies}
\label{subsec:nb_buddies}

When an event is detected, \system looks for three buddies to validate whether
the event is indeed a hijack. However, we do not always have three buddies
available. Thus, we now look at the impact of using fewer buddies on the
precision and recall of \system. Validating with more buddies is expected to
decrease the recall while improving the precision.

\begin{table}
	\centering
	\begin{tabular}{lcc}\toprule
        \textbf{Validation stage} & \textbf{Precision} & \textbf{Recall} \\\midrule
        With 3 buddies & 93.77\% & 94.26\% \\
		With 2 buddies & 92.19\% & 94.18\% \\
		With 1 buddy & 88.31\% & 94.24\% \\\midrule
		No validation & 6.11\% & 96.44\% \\\bottomrule
	\end{tabular}
	\caption{Precision and recall values for a different number of buddies that
	have to agree to report a hijack.}
	\label{tab:baseline}
\end{table}

\myitem{Setup.}
We run \system for all events in our dataset with a 3\,ms change detection
threshold and two validation steps; we compute the resulting precision and
recall values depending on the number of buddies used for the validation.

\myitem{Results.}
\cref{tab:baseline} summarizes the precision and recall values depending on
how many buddies are used for the validation.

Similarly, as in \cref{subsec:num_validations}, the recall is only marginally
affected by the number of available buddies. If the event is indeed a hijack,
buddies often validate this correctly (\ie they do not wrongly discard the
event). Interestingly, the recall is only slightly better \emph{without
validation at all} and does not reach 100\%. This is because some hijack events
induce RTT changes that are too small to be detected, given our detection
threshold. Conversely, more buddies logically yield better precision (\ie fewer
false positives).


\myitem{Conclusion.}
\system achieves good precision and recall (around 90\%) even with only one
buddy for the validation. Using more buddies results in a slight increase in
precision (a few~\%) for a negligible decrease in recall.

\subsection{Classification delay}
\label{sec:speed}

\system is meant to detect hijacks at runtime. Thus, it is important to quantify
how fast potential hijacks are detected, validated, and reported to the
operator. The time required for a hijack to propagate through the network is
beyond our control and independent of \system. The delay from detecting an event
in the combined RTT signal until the event is reported (or not) by \system is
more relevant. The delay is affected by the type of the observed event and the
number of buddies and validation steps. Intuitively, it increases with a higher
number of performed validation steps.

\myitem{Setup.}
We run \system for all events in our dataset with a 3\,ms change detection
threshold and one, two, or three validation repetitions; we measure the time
delay from detecting a change in the RTT signal until the final event
classification by \system.

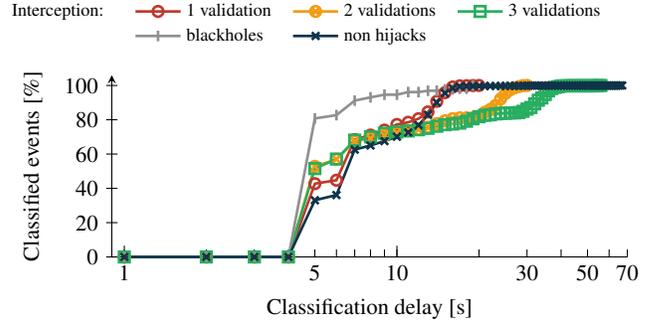
\begin{figure}
\centering

\begin{tikzpicture}[every node/.style={scale=0.85}]

  \newcommand\PlotWidth{240pt}
  \newcommand\PlotHorSep{46pt}
  \newcommand\PlotHeight{4cm}
  \newcommand\PlotVerSep{0cm}

  \begin{groupplot} [
    group style={
      group size=1 by 1,
      y descriptions at=edge left,
      horizontal sep=\PlotHorSep,
      vertical sep=\PlotVerSep,
      group name=weight group,
    },
    axis y line = left,
    ]

    \nextgroupplot[
      height=\PlotHeight,
      width=\PlotWidth,
      ylabel = {Classified events [\%]},
      axis x line* = bottom,
      xlabel = {Classification delay [s]},
      xmode=log,
      xmin = 0.9,
      xmax = 70,
      clip=false,
      xticklabels = {},
      ymax = 106,
      ytick = {0, 20, 40, 60, 80, 100},
      yticklabels = {0, 20, 40, 60, 80, 100},
      legend cell align=left,
      legend columns=3,
      legend style={
        draw=none,
        fill=none,
        font=\footnotesize,
        anchor=north,
        at={(0.5, 1.45)},
        /tikz/every even column/.append style={column sep=6pt},
      },
    ]

    \foreach \t in {1,5,10,30,50,70}{
    \edef\temp{\noexpand\node[below, text depth=0.25ex] at (axis cs: \t,0) {\t};}
    \temp
    }

    \addplot+[mark=o, cRed, solid] table[x=x, y=y, col sep=comma] {tikz/interception_1.csv};
    \addlegendentry{1 validation}

    \addplot+[mark=otimes, cYellow, solid] table[x=x, y=y, col sep=comma] {tikz/interception_2.csv};
    \addlegendentry{2 validations}

    \addplot+[mark=square, cGreen, solid] table[x=x, y=y, col sep=comma] {tikz/interception_3.csv};
    \addlegendentry{3 validations}

    \addplot+[mark=|, cGray, solid] table[x=x, y=y, col sep=comma] {tikz/bh.csv};
    \addlegendentry{blackholes}

    \addplot+[mark=x, cDarkBlue, solid] table[x=x, y=y, col sep=comma] {tikz/other.csv};
    \addlegendentry{non hijacks}

  \end{groupplot}

  \draw (0pt, 93pt) node[left] {\footnotesize Interception:};
\end{tikzpicture}

\caption{Delay between the detected change in the RTT signal and the
classification of the event by \system. Most events are classified within the
best possible delay (around 5\,s); the rest are delayed by the lack of RTT
samples from active buddies to perform the validation step(s).}

\label{fig:delay}

\end{figure}

\myitem{Results.}
\cref{fig:delay} shows the CDF of the classification delays for different
events and validation repetitions. These delays are in the scale of seconds
since \system performs the change detection process once per second. The first
observation is that the detection delay is at least 5\,s, which comes from
\system's design decision of waiting for four consecutive change detections
before starting the validation (\cref{sec:det_cpd}).

The blackhole events are the easiest to detect, with more than 80\%
classification completed after five seconds. Since no validation step is
performed for these events, the number of validation repetitions naturally has
no effect. For the interception events, around 50\% are classified after 5\,s;
that corresponds to cases where \system could quickly find buddy prefixes
carrying enough traffic for the validation step(s). Cases with higher delays are
unavoidable given the data-driven nature of \system: when there is no traffic,
there is no data to work with; one must wait for the data. With two or three
validation steps, around 80\% of interception events are classified within
20\,s. The ``non hijack'' events represent cases where \system initially
triggers, initiates the validation, and classifies the event as normal routing
changes.

Finally, the expected trend when varying the number of validations is most
sensible for the tail: for about 70\% of events, the classification delay varies
by less than 1\,s.

\myitem{Conclusion.}
By design, the classification and reporting of an event by \system takes at
least 5\,s. This choice limits the validation step from excessive triggering due
to random noise in the RTT signals. Most events are classified within 5\,s and
80\% to 100\% of them within 20\,s (depending on the event type and the number
of validation repetitions). We argue that this is a reasonable delay for
detecting hijacks that would otherwise remain unnoticed.

If faster detection is desired, one may adapt the design by
\textit{(i)}\,reducing the size of \system's long-term window and/or
\textit{(ii)}\,increasing the detection frequency (limited by the computational
load -- \cref{sec:benchmark}). However, this will only help in cases where
active buddy prefixes are available for validation. As discussed above, in our
dataset, around 50\% of interception event classifications are delayed by the
lack of available RTT samples from buddies.

\subsection{Computational overhead}
\label{sec:benchmark}

Running \system requires performing three types of computations:
\textit{(i)}\,extract the minimum RTT values from the combined signals,
\textit{(ii)}\,compare the short- and long-term minimum to detect changes, and
\textit{(iii)}\,perform the statistical tests to compare the RTT samples between
buddies. These computations must be sufficiently fast to run online; in
particular, they must scale to the amount of traffic that \system would have to
monitor in a large AS.

\myitem{Setup.}
Our implementation of \system is written in Python, similar to common software
of programmable control-plane devices. We time our implementation execution for
three classes of traffic load, quantified as the average number of RTT samples
to process per second.
\begin{itemize}[nosep,leftmargin=*]
	\item low -- 600 RTT sample/s
	\item medium -- 3000 RTT sample/s
	\item high -- 5000 RTT sample/s
\end{itemize}
We measure the timing of the operations mentioned above, labeled as
\textbf{minimum}, \textbf{change}, and \textbf{validation}, respectively.

\myitem{Results.}
\cref{tab:benchmark} summarizes the results for each operation and traffic
load. All times are in \us.

As expected, the computation of \textbf{minimum} is the one most affected by the
increasing traffic load, as there are more signals to track, thus, more
operations to perform. However, even under high load, this always took less than
250\us in our experiments. The computation is efficient as \system can compute
the minimum RTT values without storing all the samples, which is an asset of its
design. Storing a small number of sample values is only required when the
validation step(s) triggers.
The \textbf{change} operation compares two minimum values; as expected, this
operation is independent of the traffic load.
Finally, the \textbf{validation} operation performs the statistical tests
comparing RTT samples among buddies. We observe that higher loads lead to higher
computation times (two to three times more between low and high loads). This is
explained by the number of statistical tests that are effectively performed:
\system always looks for three buddies in the validation step. However, with low
traffic, there are not always enough active buddies available, thus resulting in
fewer tests being performed and lower computation times.

\myitem{Conclusion.}
As expected, the traffic load severely impacts only the \textbf{minimum}
operation. However, it takes less than 250\us even for a (high) load of about
5000 RTT samples per second. The validation step is the most computationally
expensive, reaching $\sim3\ms$. Altogether, the total computation time is in the
order of \ms; in comparison, \system runs every second, three orders of
magnitude higher. Thus, we conclude that the computational overhead of \system
is limited and, in particular, it is not a limitation for running the system
online, even under high traffic loads.

\begin{table}
	\begin{center}
	\scalebox{0.85}{
	\begin{tabular}{@{}l@{\qquad} l@{\qquad} c@{\quad}c@{\quad}c@{}}
		\textbf{operation} 	&	 \textbf{load} 	&	 \textbf{min} 	&	 \textbf{median} 	&	 \textbf{max} \\ \toprule 
		 	&	 high 	&	0.71	&	44.1	&	 244.37 \\
		\textbf{minimum} 	&	 medium 	&	0.71	&	26.7	&	 304.69 \\
		 	&	 low 	&	0.71	&	2.62	&	 111.10 \\ \midrule
		 	&	 high 	&	0.23	&	0.71	&	 16.45 \\
		 \textbf{change} 	&	 medium 	&	0.235	&	0.95	&	 16.68 \\
		 	&	 low 	&	0.23	&	0.95	&	 26.22 \\ \midrule
		 	&	 high 	&	1044.98	&	1152.75	&	 3357.64 \\
		 \textbf{validation} 	&	 medium 	&	382.18	&	1146.07	&	 2597.33 \\
		 	&	 low 	&	347.85	&	459.19	&	 2586.60 \\
	\end{tabular}
	} \caption{Only the time to find the short- and long-term \textbf{minimum}
	is highly influenced by the load (median values). The \textbf{validation}
	operation takes the longest. All values in \us.}
	\label{tab:benchmark}
	\end{center}
\end{table}

\subsection{Availability of buddies}
\label{sec:deployment}

\system relies on the availability of buddies (see \cref{sec:buddies}) to
reliably identify hijacks from the RTT signals. For each prefix, \system
searches for three buddies during its validation step. As shown in
\cref{subsec:nb_buddies}, the detection accuracy remains good even with only one
buddy available, but the precision drops when there is none. Hence, to evaluate
the usability of \system in an actual deployment, it is crucial to assess how
likely prefixes are to have buddies in the real Internet.

Assessing whether prefixes are buddies in the data plane is challenging. We can,
however, efficiently assess buddyhood in the control plane by parsing BGP
announcements. In most cases, prefixes will be buddies in the data plane when
there are buddies in the control plane, as ASes do not have per-prefix routing
policies implemented (\eg static routes). In \cref{sec:buddies}, we explain that
\system splits its prefix space into /24 prefixes for its buddy assumptions (\ie
to match the smallest possible BGP hijack). We can only make statements about
the actual prefix sizes in an analysis based on control-plane data. Therefore,
the following evaluation considers the prefix sizes observed in BGP
advertisements rather than /24.

\myitem{Setup.}
We use BGPstream \cite{bgpstream} to collect recent RIB dump available from 44
RIPE RIS and RouteViews BGP collectors. Each collector gathers and reports on
the best routes from multiple collaborative ASes (BGP peers), so-called
monitors. We define prefixes as buddies from one monitor's point of view if they
originate from the same AS and their announcements contain the same AS path. For
each prefix, we count the number of such buddy prefixes seen by the monitor.
Since all RIRs deploy RIPE RIS and RouteViews monitors, we expect them to
capture the level of buddyhood in a large part of the Internet. Note that two
buddies might share part of their prefix size, \eg if one AS advertises a /16
prefix and a /20 prefix which is a sub-prefix of the /16 one.

\begin{figure}
	\centering
	\includegraphics[width=\columnwidth]{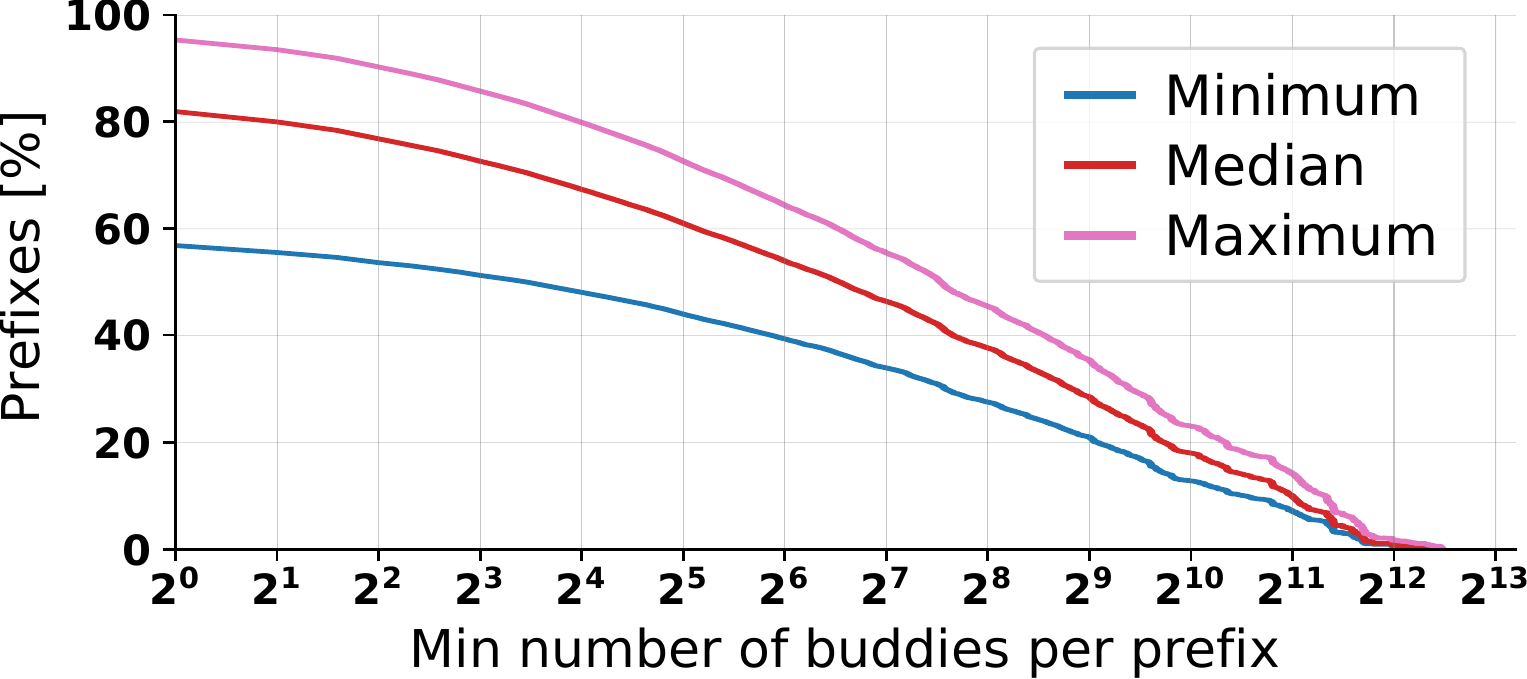}
	\caption{Number of buddies per prefix seen by 44 RIPE RIS and RouteViews BGP
	collectors. The different lines indicate the ``minimum'', ``median'', and
	``maximum'' numbers of buddies seen by the various monitors.}
	\label{fig:ideal}
\end{figure}

\myitem{Results.}
\cref{fig:ideal} shows a complementary CDF of the number of buddies found per
prefix. The different lines show the minimum, median, and maximum numbers of
buddies seen by the various monitors. In the ``worst-case'' (\ie min. line),
about 55\% of prefixes have at least three buddies, and about 57\% have at least
one. We observe a significant variation between the monitors: the median monitor
sees about 79\% of prefixes with at least three buddies; the value goes to 92\%
for the ``best'' monitor.

\myitem{Conclusion.}
Our analysis shows that a majority of prefixes have at least three buddies.
However, the different monitors see a significant variation in the buddies.
Thus, one cannot generalize: buddyhood strongly varies between ASes.

Furthermore, there remains a large portion (between 5 to 45\%) of prefixes for
which the monitors do not see buddies. ``Partial buddies'' (only routed in the
same way in parts of the Internet) can also be helpful for \system; the partial
buddyhood is significantly larger (not shown in the plot). But for prefixes
without buddies, \system cannot reliably validate RTT events as interception
hijacks. This is an inherent limitation of \system's data-driven design.

\pagebreak

\section{Discussion}
\label{sec:discussion}

This section discusses advanced aspects of \system regarding deployability and
performance under special cases.

\myitem{Does \system replace existing systems?} No, we aim not to replace
existing RPKI and monitor-based solutions but to complement them. For instance,
\system cannot detect attacks where the RTT remains unchanged or the hijacker
intercepts \emph{all} the IP prefixes in the same buddy group. In these cases,
one has to resort to monitor-based solutions. \system also does not prevent
hijacks, making crypto-based solutions still beneficial (yet challenging to
deploy).

\myitem{Can \system detect pre-existing hijacks?} 
If an operator suspects that a specific IP prefix might be hijacked (\eg based
on insights from the control plane), we can manually trigger the validation
step, which will compare samples between buddy prefixes and verify the hijack.

\myitem{Can a hijacker hide from \system?} A hijacker may try to evade detection
by tweaking the RTT signal or generating short-time hijacks. \textit{RTT
Signal:} It is hard (or impossible in some cases) to evade detection by merely
tweaking the RTT signal. First, if the hijacking path has a higher RTT than the
legitimate one, there is no way a hijacker can beat the speed of light to
pretend the RTT is unchanged. Second, a hijacker that induces a lower RTT may
hold packets back but guessing the legitimate RTT distribution would be out of
reach since the hijacker has no visibility into those distributions. More
precisely, usually, the attacker does not know what the RTT before the hijack
was, as the attacker is (by definition) \emph{not} on the legitimate path.
\textit{Short-time hijacks:} The attack should last long enough to propagate the
announcement to the victim AS. In the meantime, the RTT signal of the
intermediate ASes would be affected and could be detected by \system.
High-frequency short-time attacks may be filtered by BGP Flap
Damping~\cite{bgp}.

\myitem{Can \system detect a hijack towards one of its prefixes for which we
currently do not observe any RTT samples?} In its current form, \system is
entirely data-driven and cannot detect hijacks for which we do not collect
corresponding data-plane signals, \ie RTTs. However, that also means that the
hijack is less severe as it does not affect any actual traffic.

\myitem{How can an operator regulate the number of alerts that \system produces?}
As we show in \cref{sec:evaluation}, the operator could, for example, increase
the change detection threshold, improving the precision and removing some of the
noise/false positives, \ie reported events often indicate a real hijack. Another
promising approach extends one of the previous discussion points. An operator
could combine \system's reports with observations from a control-plane-based
detection system. For both systems, we can filter out reports with low
confidence unless both systems raise them. For example, if \system reports a
hijack validated with a single buddy prefix only.

\myitem{What happens with false buddy prefixes, \ie they do not follow the same
forwarding path?} So far, we assumed that \system's validation step discards
results if not all performed tests agree. However, our analysis in
\cref{sec:deployment} shows that large buddy groups are common. As a result,
\system could base its decision on a majority vote instead, which will filter
out isolated, false buddies.

\myitem{Does buddyness hold within a single AS?} We show in
\cref{sec:deployment} that many IP prefixes propagate along the same AS path
over the Internet. Many intra-domain Traffic-Engineering mechanisms, \eg
OSPF~\cite{ospf}, RIP~\cite{rip}, MPLS auto-bw~\cite{mpls}, split traffic across
different physical paths using hash-based mechanisms. While such mechanisms may
induce different RTTs for distinct flows within a single IP
prefix~\cite{aws-te}, multiple prefixes exhibit the same distribution and thus
would be correctly handled by \system. We believe that \system should be
complemented with a buddy-tracking system (left as future work) that identifies
IP prefixes that exhibit a similar RTT distribution and time-correlated RTT
changes. This paper focuses on understanding whether one could extract and
combine RTT signals assuming buddies exist.

\myitem{Does \system need to be deployed in multiple ASes to work correctly?}
No, it is enough to deploy \system in a single AS. It is only designed to report
hijacks that affect prefixes originated by this AS. \system extracts all the
required information locally. For example, RTT samples from flows that start/end
in the local AS or BGP data from local routers (\ie to build \emph{combined
signals}). However, aggregating reports from multiple \system instances deployed
in different ASes could help reduce false positives or detect large-scale
hijacks that might affect prefixes from multiple ASes at once.

\section{Conclusion}
\label{sec:conclusion}

In this paper, we show that one can leverage data-plane signals to mitigate a
long-standing security problem of inter-domain routing: detecting BGP hijacks.
We believe it is a first step towards the general question of classifying
large-scale Internet events by fully integrating data- and control-plane
analysis methods. It opens and calls for further investigation, \eg applications
of ML to classify the different signals.

\clearpage

\bibliographystyle{plain}
\bibliography{references}

\clearpage

\appendix

\section{The Stealthy Hijacker}\label{app:stealthy-hijacker}

In the following sections, we explain what we understand under ``stealthy
hijacks'' (\cref{sec:stealthy-hijacker}), how our Internet simulator
works (\cref{app:Simulator}), and how we use the simulator to evaluate the
impact of stealthy hijacks (\cref{sec:StealthyGen}).

\subsection{Generating stealthy hijacks}
\label{sec:stealthy-hijacker}

Recently, there have been insights into techniques that smarter hijackers can
leverage to evade traditional detection systems. SICO~\cite{birge2019sico}
successfully managed to leverage BGP Community Attributes to surgically limit
the propagation of the attack to only the intended target and networks along the
path. Smith et Al.~\cite{smith2020withdrawing} measured the impact of BGP
poisoning, a technique that hijackers can also use to shape traffic paths, and
found that only a few ASes implement filters against it. Testart et
Al.~\cite{testart2019profiling} used Machine learning techniques to identify
networks of serial hijacking activity that were previously unnoticed. This
concept of stealthy hijackers dates back even in 2009 with the exploit of
``Fakeroute'' tools \cite{mcarthur2009stealthy}; systems can lie to data-plane
probes to distinguish hijacks from other routing problems, \eg by faking the
hijack signal with the signal of normal routing failures.

In this work, we develop a smart hijacker to hide from control-plane monitors
through BGP poisoning and path shaping. Our insight is that a hijacker, who
knows that ISPs provide routing information to the BGP collectors, can shape its
hijack announcements, making them invisible to those collectors and, thus, the
monitors. Succeeding in such an attack could render most, if not all, current
control-plane monitoring solutions ineffective. This includes tools like
Artemis~\cite{artemis}, Argus~\cite{argus}, BuddyGuard~\cite{buddyguard}, etc.

Through the use of simulations, we measured the impact of our stealthy hijacker.
As depicted in \cref{fig:scale}, compared to a normal hijacker whose attack
becomes visible and the number of poisoned networks negligible, our hijacker can
evade detection even as we scale the number of monitors.

\subsection{Our Internet simulator}
\label{app:Simulator}

We used a well-tested network simulator developed in a previous literature study
to implement our hijacker. In this simulator, benign networks announce only the
prefixes they originate. Once an AS receives a new prefix announcement from its
neighbors, a) it updates its routing table with the new route towards the
prefix, then b) recalculates the best path towards that prefix, and finally, c)
propagates the new best route (if any) towards a choice of its neighbors.

Each AS consults its routing table to select the best path and compares the
newly received route against the current best route for that received prefix.
First, the one with the highest local preference attribute will be preferred,
then the one with the shortest AS path length, and finally, the one with the
lowest tie-breaking attribute. Since ISPs are commonly commercial businesses, we
model networks to prefer the paths from their customers, as they generate
profit, over the paths from their peers that offer no revenue, over the paths
from their providers that generate costs~\cite{gao2001stable}. Each AS refers to
a random tiebreaker for paths of equal preference and length to choose the best
route. In the real Internet, networks usually configure such tie-breaking
mechanisms based on their internal organization and inbound traffic engineering
goals~\cite{caesar2005bgp}. In the simulator, though, since we have no
visibility of those attributes, we randomly configure them during the topology
setup.

Once an AS updates its best path, it propagates it to the rest of its neighbors.
We model path propagation according to the Gao-Rexford
conditions~\cite{gao2001stable}. Since ASes usually seek to maximize their
profit, we model paths received from customers to propagate to every neighbor.
In contrast, paths received from providers or peers will propagate only to
customer nodes.

Under every circumstance, ASes will drop BGP advertisements containing their ASN
in the AS-path attribute, reflecting the BGP loop avoidance mechanism.
Furthermore, for backup reasons, they will update their routing entries with the
new path even if that is not the new best one. Lastly, they will append their
ASN as the last AS in the path before propagating the route further.

Hijacker networks may craft routes from scratch, modify any advertisement they
receive, and advertise routes to any of their neighbors. Their goal is not to
propagate valid routes and generate profit but rather to hijack traffic going
back to the victim's prefix.

To model the Internet topology, we use CAIDA's AS relationship
datasets~\cite{CaidaASrelations} which are collected from the RIPE RIS and
Routeviews BGP collectors. CAIDA processes those data through a multi-step
methodology to produce a sanitized view that provides a well-designed
representation of the Internet widely used in related work. Each node represents
an ASN with interconnections observed in the BGP routes. The relations of those
connections are categorized as i) peer-to-peer, ii) customer-to-provider, or
iii) provider-to-customer (if seen in the opposite direction).

We enhance those data with additional peering interconnections from networks at
Internet exchange points (IXPs). We compared a sample of the simulated paths
with the corresponding real Internet paths observed from a BGP looking glass.
Our results indicate that the simulated paths are longer, in comparison, due to
missing interconnections at local IXPs. As such, we use CAIDA's IXP
datasets~\cite{CAIDAIXP} to include possible cases of missing peering
interconnections. For each IXP, we first identify its members and then assign
them to a peer-to-peer relation if these connections were missing from our
simulator. If a different relation existed between the members in question, we
would maintain the old ones inferred by CAIDA.

\pagebreak

Furthermore, we mark in our simulator the data providers of BGP
Stream~\cite{Streamcollectors}. Those are the ASes peering with the collection
points providing snapshots of their best paths to the public BGP Stream service.
We identify those networks by analyzing a day of BGP dumps from the BGPstream
service. By extracting the ``peer\_ASN'' attribute from the update dumps, we
ended up with 408 unique peers, out of which 320 actively reported paths within
5 minutes. Since the ``not-so-active'' peers cannot be trusted to report hijacks
promptly, we filter them out. We refer to the remaining peers as BGP monitoring
locations or monitors in short. Such as the networks M1 and M2 depicted in
\cref{fig:hijack}.

\subsection{Generating stealthy hijacks}
\label{sec:StealthyGen}

We test the stealthy hijacker using the Internet simulator. We randomly pick one
victim and one hijacker network in each experiment. Then the hijacker launches a
same-prefix attack for one of the prefixes announced by the victim. As in
\cref{fig:hijack}, this will cause part of the Internet to prefer the hijacker
(poisoned networks) while the rest will still pick the victim. This is following
the hijacker's goals, who does not seek to maximize the scale of the attack but
rather launch it in such a way as to make it undetectable by the monitors. It
also benefits the hijacker as attacks of a smaller scale yield less traffic, are
easier to intercept back, and are less likely to be noticed by the victim.

When picking a new victim node, we verify that it is not disconnected, \ie that
the victims' announcement propagates almost the entire Internet. Furthermore,
when selecting the hijacker, we verify that a) it has at least one available
route back to the victim, b) it has at least two neighbors, and c) it is neither
a monitor network nor a direct neighbor of the victim as we assume that those
networks can be trusted. We require the hijacker to have at least two neighbors,
one to launch the hijack from and the other to intercept the traffic back to the
correct origin. \cref{fig:hijack} depicts such an example. Assume that Bob and
Alice are the randomly picked hijacker-victim pair. Alice needs to craft a
stealthy hijack announcement that is not visible to the monitor. She does that
based on the following methodology. \\
a) She identifies the neighbors that advertise a route towards Bob's prefix.
These are the neighbors N1 and N2. For the stealthy hijack to be successful as
an interception attack, one of those routes needs to be maintained after the
attack. \\
b) She fetches the best BGP routes towards the prefix of Bob as advertised by M1
and M2 to the BGP collector. This information is publicly available via the
monitors. \\
c) She fetches the same information but this time for her own prefix. Alice
announces a different prefix from her own address space to each of her neighbors
(P1, N1, N2). After those advertisements propagate, Alice uses the collector to
see how the monitors observe those routes. Alice will use the information
extracted from steps b) and c) to craft a smart announcement individually to
each of her neighbors. Note that so far, there is nothing malicious in Alice's
behavior. \\
d) Using step b), how the monitors observe Bob's prefix, and step c), how the
monitors observe Alice's prefixes, Alice estimates the ``dangerous'' monitors
who are likely to observe the attack. Since Alice is unfamiliar with the local
preferences of other networks, she does this estimation by comparing the length
of her routes versus the length of Bob's route as observed by each monitor. If
Alice's route is shorter or equal to Bob's route for a monitor, then, because in
BGP, shorter paths are preferred, the monitor is marked as dangerous. Note that
Alice does this calculation independently for each of her neighbors. For
example, for the crafted route \{~Alice, P1, Bob~\} that Alice would like to
announce to her neighbor N2, Alice marks M2 as a dangerous monitor since her
advertisement \{~N2, Alice, P1, Bob~\} is shorter than the valid route \{~N2,
M1, N1, P2, Bob~\}, as seen by M2. For her neighbor N1, Alice would perform the
same dangerous monitor calculation. However, as she observes that announcing a
crafted route to N1 will break her valid route back to Bob, she will refrain
from making that announcement. \\
e) After discovering the dangerous monitors, Alice will announce a hijack route
for Bob's prefix that is not reportable by the monitors. Since Alice only
discovered a single monitor, she poisons her announcement with the ASN of that
monitor, \ie \{ Alice, M2, P1, Bob \}. When M2 receives the poisoned
advertisement and sees its ASN inside, it will drop it due to BGP's loop
avoidance algorithm. As such, since no monitor will report the announcement, it
will not be considered by control-plane monitoring solutions.

In our simulator, we have marked a total of 320 monitors. Including all of those
monitors in the crafted announcement will create an unrealistic route. Instead,
when multiple monitors observe the hijack advertisement, we poison with the ASNs
of the most common ASes observed between the hijacker and the monitors, either
to block the propagation of the hijack to the monitors or to make the hijack
less preferable. Since multiple routes exist, blocking the best route may cause
the hijack to propagate to the monitor via a different route. Given that the
hijacker has a reasonable amount of time, it could shape its announcement to
block every path. However, for our simulations, if Alice fails on the first try,
we mark the stealthy hijack as a failure, as Bob could potentially notice it.
The purpose of this paper is not to find the best way for the hijacker to hijack
the victim stealthily but to verify in simulations that such attacks are
possible. As mentioned in previous work~\cite{goldberg2010secure}, identifying
the best way to hijack the victim is an NP-hard problem.

In total, we perform 2000 simulations randomly picking pairs of Alice and Bob.
Alice tries to stealthily hijack Bob in each simulation using the methodology
described above. Since, in this methodology, Alice could craft routes of
arbitrary length, we limit her announced path length to type-4
hijacks~\cite{artemis}. Crafted type-4 hijack are of the form \{Alice, ASA, ASB,
ASC, Bob\} where Alice and Bob are the corresponding hijacker and hijacked AS,
and the ASNs in between, hops that Alice can modify to craft the stealthy
announcement. Alice announces the correct origin, Bob, so her advertisements are
RPKI valid. Furthermore, we perform a baseline comparison to the stealthy hijack
in each simulation. We configure Alice to announce a naive type-4 hijack to each
of her neighbors using path prepending, \ie \{ Alice, Bob, Bob, Bob, Bob\}. Last
but not least, we verify which impact the number of monitors has on the
visibility of our stealthy hijacker. We scale the number of monitoring ASes up
to a factor of five (320x5 = 1600 monitors) following a distribution similar to
that of BGPStream. Using~\cite{caida_asrank,caida_asrank_new}, we extract the AS
ranks of the monitors of BGPStream. Then, for each monitor AS, we add a new
monitor from the closest available AS with a ranking similar to the rank of the
existing monitor.

\cref{fig:scale} depicts the stealthiness of the naive hijacker compared to
our stealthy hijacker as we scale the number of monitors. For the results to be
comparable, we run 2000 simulations using the same randomization seed for each
scale. Furthermore, as we care only for the additional traffic that Alice
attracts after launching the hijack, we ignore the networks that were using
Alice for their best route towards Bob, to begin with, \ie before the hijack.
From \cref{fig:scale}, we observe the following:
\begin{enumerate}
	\item Naive hijackers are likely to generate stealthy attacks based on the
	relevant positions of the hijacker, the victim, and the monitor networks. As
	we observe, for the base number of monitors, $\approx14\% (280/2000)$ of the
	hijacks were invisible, attracting traffic from $\approx2.4\% (1580/65724)$ of
	our typology's networks. However, as we scale the number of monitors, the
	naive hijack impact soon becomes negligible, affecting $<1\%$ of the
	networks while being invisible.
	\item Our smart hijacking methodology for the base number of monitors can
	more consistently generate stealthy hijacks with a $\approx31\% (620/2000)$
	success rate attracting traffic of $\approx5.5\% (3600/65724)$ of the Internet.
	As we scale the number of monitors, we still observe a $\approx24\% (480/2000)$
	success rate, with $\approx~3.8\% (2500/65724)$ of the networks redirecting
	their traffic to Alice.
\end{enumerate}

\textbf{Sum-up:} Existing monitoring solutions~\cite{artemis, argus, buddyguard,
testart2019profiling, lad2006phas, chi2008cyclops, yan2009bgpmon} that operate
the core of their detection module in the control plane will not identify a
smart hijacker that succeeds in a stealthy attack against the monitors. Existing
data-plane monitoring solutions~\cite{ispy, hu2007accurate} relying on
connectivity loss or fingerprint extraction will fail to identify the smart
hijacker since we retain a route back to the victim. Unless very sensitive,
techniques relying on traceroute hops~\cite{zheng2007light} might be outsmarted
by an intelligent hijacker able to produce fake IP hops or distinguish the
hijack as a common routing problem. \\

\end{document}